\documentclass[prd,preprintnumbers,twocolumn,amsmath,nofootinbib,amssymb]{revtex4}
\usepackage{graphicx,color,dcolumn,booktabs,bm}
\usepackage{longtable,lscape}
\usepackage{txfonts}
\usepackage{overpic}
\usepackage{amssymb}
\usepackage{epstopdf}
\usepackage{indentfirst}
\usepackage{feynmf}   
\usepackage{slashed}  
\usepackage{cases}
\usepackage{color}
\usepackage{float}
\usepackage{multirow}
\usepackage{ulem}
\usepackage{graphicx,color,dcolumn,booktabs,bm}
\usepackage{epsfig,dsfont,amssymb,amsmath,amsfonts,amsbsy,mathrsfs}

\graphicspath{{Figures/}} %

\usepackage{hyperref}
\hypersetup{colorlinks,citecolor=blue,anchorcolor=red,menucolor=red, linkcolor=red,filecolor=red,runcolor=red,urlcolor=blue,frenchlinks=red}

\makeatletter
\@addtoreset{equation}{section}
\makeatother

\allowdisplaybreaks

\begin{document}

\title{Investigating new type of doubly-charmed molecular tetraquarks composed of the charmed meson in $H$-doublet and the charmed meson in $T$-doublet}
\author{Fu-Lai Wang$^{1,2}$}
\email{wangfl2016@lzu.edu.cn}
\author{Xiang Liu$^{1,2,3}$\footnote{Corresponding author}}
\email{xiangliu@lzu.edu.cn}
\affiliation{$^1$School of Physical Science and Technology, Lanzhou University, Lanzhou 730000, China\\
$^2$Research Center for Hadron and CSR Physics, Lanzhou University and Institute of Modern Physics of CAS, Lanzhou 730000, China\\
$^3$Lanzhou Center for Theoretical Physics, Key Laboratory of Theoretical Physics of Gansu Province, and Frontiers Science Center for Rare Isotopes, Lanzhou University, Lanzhou 730000, China}

\begin{abstract}
Stimulated by the newly reported doubly-charmed tetraquark state $T_{cc}^+$ by LHCb, we carry out a systematic investigation of the $S$-wave interactions between the charmed meson ($D,\,D^{*}$) in $H$-doublet and the charmed meson ($D_{1},\,D_{2}^{*}$) in $T$-doublet by adopting the one-boson-exchange model. Both the $S$-$D$ wave mixing effect and the coupled channel effect are taken into account. By performing a quantitative calculation, we suggest that the $S$-wave $D^{*} D_{1}$ states with $I(J^{P})=0(0^{-},\,1^{-})$ and the $S$-wave $D^{*}D_{2}^{*}$ state with $I(J^{P})=0(1^{-})$ should be viewed as the most promising candidates of the doubly-charmed molecular tetraquark states, and the $S$-wave $DD_{1}$ state with $I(J^{P})=0(1^{-})$, the $S$-wave $DD_{2}^{*}$ state with $I(J^{P})=0(2^{-})$, and the $S$-wave $D^{*}D_{2}^{*}$ state with $I(J^{P})=0(2^{-})$ are the possible doubly-charmed molecular tetraquark candidates. With the accumulation of experimental data at Run III and after High-Luminosity-LHC upgrade, these predicted doubly-charmed molecular tetraquark states can be accessible at LHCb in the near future.
\end{abstract}

\maketitle

\section{Introduction}\label{sec1}

The study of the exotic hadronic states, which cannot fit into the conventional meson and baryon \cite{GellMann:1964nj,Zweig:1981pd}, has become one of the central topics at the precision frontier of the hadron physics. As an important part of the hadron spectroscopy, these exotic hadronic states include the hybrid mesons, glueballs, multiquark states, and so on \cite{Chen:2016qju,Liu:2019zoy}. Studying the exotic hadronic states can provide the valuable hints to shed light on the non-perturbative behavior of the quantum chromodynamics (QCD). Since the observation of the $X(3872)$ in 2003 \cite{Choi:2003ue}, a series of charmonium-like $XYZ$ states and several $P_c$ states 
had been discovered experimentally \cite{Chen:2016qju,Liu:2019zoy,Olsen:2017bmm,Guo:2017jvc,Liu:2013waa,Hosaka:2016pey,Brambilla:2019esw}. Since the masses of these $X/Y/Z/P_c$ states are close to the corresponding thresholds of two hadrons, the explanations of the hidden-charm molecular tetraquark and pentaquark states were extensively proposed and studied in the past decades \cite{Chen:2016qju,Liu:2019zoy}.

Very recently, the LHCb Collaboration reported an important observation of the doubly-charmed tetraquark state $T_{cc}^+$ in the $D^0D^0\pi^+$ invariant mass spectrum via the proton-proton collisions \cite{Tcc:talk}. The Breit-Wigner parameters of this doubly-charmed tetraquark state are $\delta m = -273\pm 61\pm 5^{+11}_{-14}~\text{keV}/c^2$ and $\Gamma = 410\pm 165\pm 43^{+18}_{-38}~\text{keV}$, where the mass difference is with respect to the $D^0D^{*+}$ threshold \cite{Tcc:talk}. We should emphasize that the observation of the doubly-charmed tetraquark $T_{cc}^+$ is a great breakthrough for the hadron physics, and it is the firstly observed doubly-charmed tetraquark state with the typical quark configuration $cc\bar{u}\bar{d}$, which can be assigned as the isoscalar $DD^*$ molecular state with $J^{P}=1^{+}$ \cite{Li:2012ss,Xu:2017tsr,Liu:2019stu,Li:2021zbw,Chen:2021vhg}.
\begin{figure}[!htbp]
\centering
\begin{tabular}{c}
\includegraphics[width=0.45\textwidth]{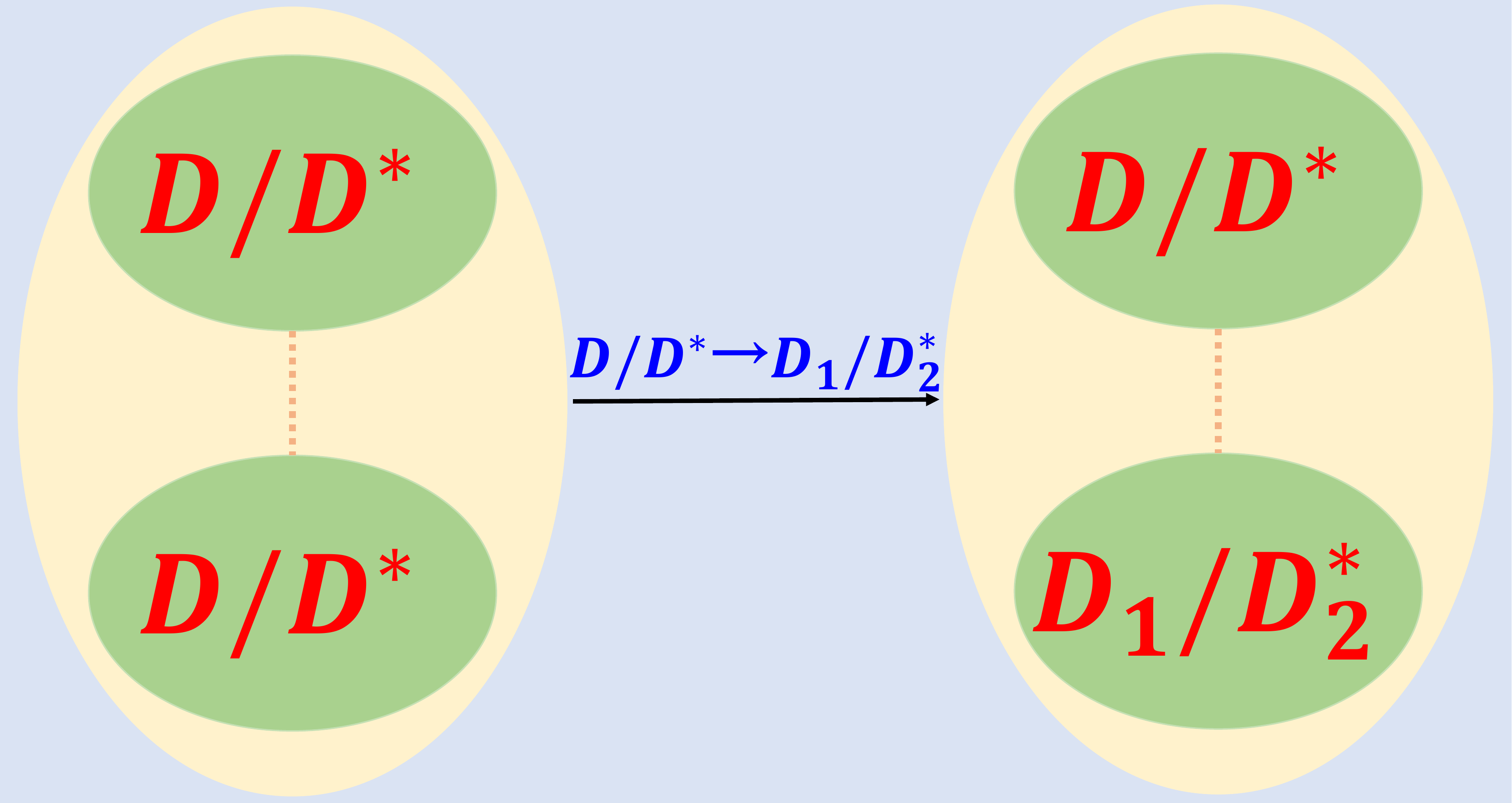}
\end{tabular}
\caption{With a simple replacement of the charmed mesons, studying the interactions between the charmed meson in $H$-doublet and the charmed meson in $T$-doublet is the natural extension of former investigation of the interactions involving two $S$-wave charmed mesons \cite{Li:2012ss,Xu:2017tsr,Liu:2019stu}.}\label{Extension}
\end{figure}

In 2003, the charmoniumlike state $X(3872)$ was observed by the Belle Collaboration \cite{Choi:2003ue}, and more theoretical groups joined the discussion of whether the $X(3872)$ can be viewed as the $D\bar{D}^*$ molecular state \cite{Wong:2003xk,Swanson:2003tb,Suzuki:2005ha,Liu:2008fh,Thomas:2008ja,Liu:2008tn,Lee:2009hy,Zhao:2014gqa,Li:2012cs,Voloshin:2003nt,Close:2003sg,Tornqvist:2004qy,He:2014nya}, which is involved in the depiction of the interactions between an $S$-wave charmed meson and an $S$-wave anti-charmed meson. Later, the experimental collaborations reported the more and more observations of charmonium-like states with higher mass \cite{Brambilla:2019esw}, which results in the discussion of whether they can also be regarded as the hidden-charm molecular tetraquark states by checking the interactions of an $S$-wave charmed meson and a $P$-wave anti-charmed meson \cite{Chen:2016qju,Liu:2019zoy}. For example, the $Z(4430)$ \cite{Choi:2007wga} can be interpreted as the $D^*\bar{D}_1^{(\prime)}$ molecular state \cite{Liu:2007bf,Liu:2008xz}, the $Y(4260)$ \cite{Aubert:2005rm} as the $D\bar D_1$ molecular state was given in Refs. \cite{Ding:2008gr,Cleven:2013mka,Wang:2013kra}, the $Y(4360)$ \cite{BaBar:2006ait} was interpreted as the $2S$ bound state in the $D^* \bar D_1$ system \cite{Close:2009ag,Close:2010wq}, and so on. If borrowing the research experience of the hidden-charm molecular tetraquark states \cite{Chen:2016qju,Liu:2019zoy}, it is time to make realistic study of the doubly-charmed molecular tetraquark candidates composed of the charmed meson in $H$-doublet and the charmed meson in $T$-doublet. Meanwhile, the study of the interactions between the charmed meson in $H$-doublet and the charmed meson in $T$-doublet is the natural extension of former study of the interactions involved two $S$-wave charmed mesons \cite{Li:2012ss,Xu:2017tsr,Liu:2019stu} (see Fig. \ref{Extension}), which has inspired the observation of the doubly-charmed $T_{cc}^+$ state to some extent \cite{Tcc:talk}. Recently, these possible doubly-charmed molecular tetraquark states were considered with the light vector meson exchange in Ref. \cite{Dong:2021bvy}.

Considering future experimental potential, we believe that the doubly-charmed molecular tetraquark states can be accessible through the proton-proton collisions. In 2018, the LHC Collaboration announced a white paper on the future physics program \cite{Bediaga:2018lhg}, Run III will be carried out with higher statistic data accumulation in near future, and the High-Luminosity-LHC will be upgraded for the physics reach. Thus, it is possible that more doubly-charmed molecular tetraquark states will be found by LHCb in near future. As an area full of opportunities and challenges, exploring various types of the doubly-charmed molecular tetraquark states should be paid more attention by both theorists and experimentalists.  

In this work, we perform a quantitative calculation of the $S$-wave interactions between the charmed meson in $H$-doublet and the charmed meson in $T$-doublet. For achieving this goal, we adopt the one-boson-exchange (OBE) model, which was extensively applied to study the hadronic molecular states \cite{Chen:2016qju,Liu:2019zoy}, to deduce the effective potentials for these discussed $S$-wave $HT$ systems. In our calculation, the $S$-$D$ wave mixing effect and the coupled channel effect are taken into account. Based on the obtained effective potentials, we may get the bound state properties of these discussed $S$-wave $HT$ systems, and further judge whether these discussed new type of the doubly-charmed molecular tetraquark states exist or not, by which we hope that these crucial information can encourage experimental colleagues to focus on these predicted doubly-charmed molecular tetraquark states in near future. In addition, we hope that the join effort from both experimentalist and theorist makes the family of the doubly-charmed molecular tetraquark states become more complete \cite{Li:2012ss,Xu:2017tsr,Liu:2019stu}.

This paper is organized as follows. In Sec. \ref{sec2}, we present the detailed deduction of the $S$-wave interactions between the charmed meson in $H$-doublet and the charmed meson in $T$-doublet by adopting the OBE model, and the corresponding bound state properties of the $S$-wave $HT$ systems will be presented in Sec. \ref{sec3}. Finally, we will give a summary in Sec. \ref{sec4}.

\section{The detailed deduction of the $HT$ interactions}\label{sec2}

In the present work, we mainly want to answer whether or not the charmed meson in $H$-doublet and the charmed meson in $T$-doublet can be bound together to form new type of the doubly-charmed molecular tetraquark states. Firstly, we study the $S$-wave effective interactions between the charmed meson in $H$-doublet and the charmed meson in $T$-doublet within the OBE model \cite{Chen:2016qju,Liu:2019zoy}, including the contribution from the $\sigma$, $\pi$, $\eta$, $\rho$, and $\omega$ exchange interactions in our concrete calculation, which is inspired by the experience of the nuclear force \cite{Chen:2016qju,Liu:2019zoy}. Here, the $\pi$, $\sigma/\eta$, and $\rho/\omega$ exchanges provide the long-range, intermediate-range, and short-range interactions, respectively.

In the following, the general procedure for deducing the OBE effective potentials in the coordinate space for these discussed $HT$ systems is presented \cite{Wang:2020dya,Wang:2019nwt,Wang:2019aoc,Wang:2020bjt,Wang:2021hql,Chen:2018pzd,Wang:2021aql}.

\subsection{The scattering amplitude}
\begin{figure}[!htbp]
\centering
\begin{tabular}{cc}
\includegraphics[width=0.20\textwidth]{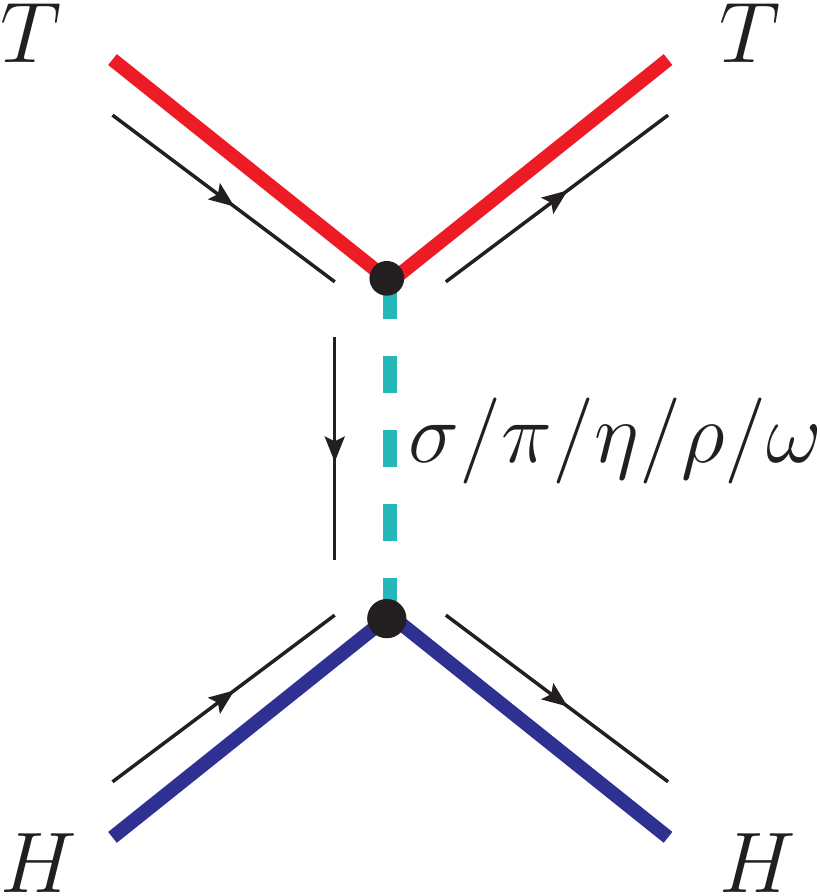}~~~~
\includegraphics[width=0.20\textwidth]{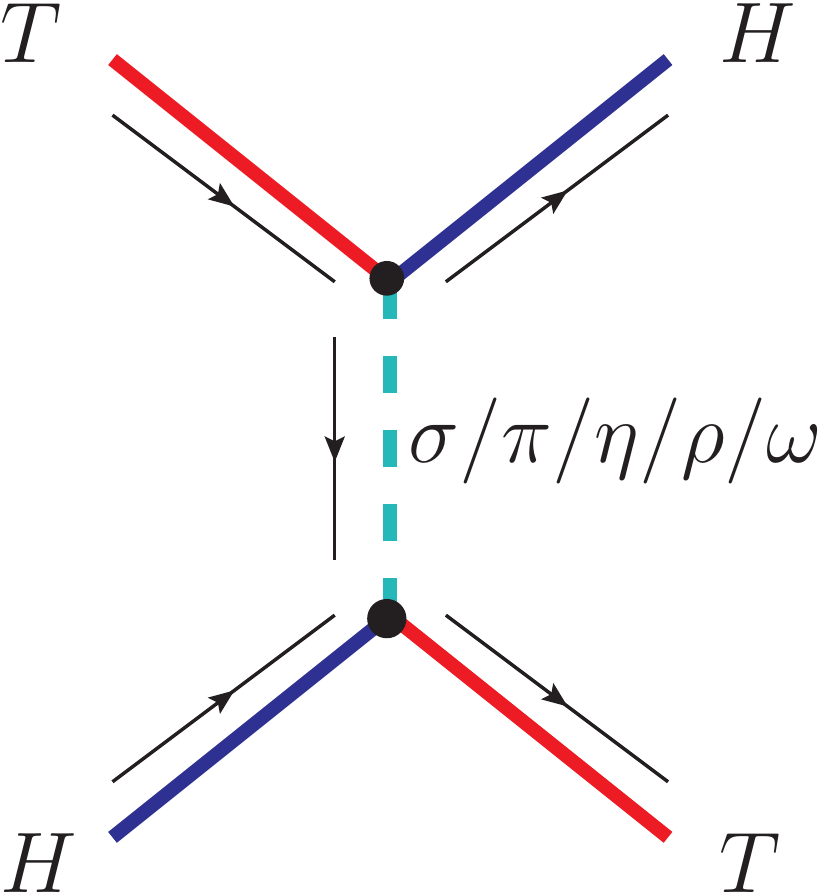}
\end{tabular}
\caption{The relevant Feynman diagrams for the scattering process $HT \to HT$. Here, $H$ and $T$ stand for the $D(D^*)$ and $D_1(D_2^*)$, respectively.}\label{fy}
\end{figure}

In the framework of the OBE model, we firstly write out the scattering amplitude $\mathcal{M}(h_1h_2\to h_3h_4)$ for the scattering process $h_1h_2\to h_3h_4$ by exchanging allowed light mesons. At the hadronic level, we usually adopt the effective Lagrangian approach to express the concrete scattering amplitude quantitatively. For the scattering process $HT \to HT$, the relevant Feynman diagrams are given in Fig.~\ref{fy}. Here, we need to construct the corresponding effective Lagrangians to describe these related interactive vertexes.

By considering the heavy quark symmetry, the chiral symmetry, and the hidden local symmetry \cite{Casalbuoni:1992gi,Casalbuoni:1996pg,Yan:1992gz,Harada:2003jx,Bando:1987br}, one constructs the effective Lagrangians relevant to the charmed mesons in the $H/T$-doublet and the light scalar, pseudoscalar, or vector mesons, where the compact effective Lagrangians can be explicitly written as \cite{Ding:2008gr}
\begin{eqnarray}\label{eq:compactlag}
{\mathcal L}_{\sigma}&=&g_{\sigma}\left\langle H^{(Q)}_a\sigma\overline{H}^{(Q)}_a\right\rangle+g^{\prime\prime}_{\sigma}\left\langle T^{(Q)\mu}_a\sigma\overline{T}^{(Q)}_{a\mu}\right\rangle\nonumber\\
&&+\frac{h^{\prime}_{\sigma}}{f_{\pi}}\left[\left\langle T^{(Q)\mu}_a\partial_{\mu}\sigma\overline{H}^{(Q)}_a\right\rangle+h.c.\right]
\end{eqnarray}
for the light scalar meson exchange,
\begin{eqnarray}\label{eq:compactlag}
{\mathcal L}_{\mathbb P}&=&ig\left\langle H^{(Q)}_b{\mathcal A}\!\!\!\slash_{ba}\gamma_5\overline{H}^{\,({Q})}_a\right\rangle+ik\left\langle T^{\,(Q)\mu}_b{\mathcal A}\!\!\!\slash_{ba}\gamma_5\overline{T}^{(Q)}_{a\mu}\right\rangle\nonumber\\
&&+\left[i\left\langle T^{(Q)\mu}_b\left(\frac{h_1}{\Lambda_{\chi}}D_{\mu}{\mathcal A}\!\!\!\slash+\frac{h_2}{\Lambda_{\chi}}D\!\!\!\!/ {\mathcal A}_{\mu}\right)_{ba}\gamma_5\overline{H}^{\,(Q)}_a\right\rangle+h.c.\right]\nonumber\\
\end{eqnarray}
for the light pseudoscalar meson exchange, and
\begin{eqnarray}\label{eq:compactlag}
{\mathcal L}_{\mathbb V}&=&\left\langle iH^{(Q)}_b\left(\beta v^{\mu}({\mathcal V}_{\mu}-\rho_{\mu})+\lambda \sigma^{\mu\nu}F_{\mu\nu}(\rho)\right)_{ba}\overline{H}^{\,(Q)}_a\right\rangle\nonumber\\
&&+\left\langle iT^{\,(Q)}_{b\lambda}\left(\beta^{\prime\prime} v^{\mu}({\mathcal V}_{\mu}-\rho_{\mu})+\lambda^{\prime\prime}\sigma^{\mu\nu}F_{\mu\nu}(\rho)\right)_{ba}\overline{T}^{(Q)\lambda}_{a}\right\rangle\nonumber\\
&&+\left[\left\langle T^{(Q)\mu}_b\left(i\zeta_1({\mathcal V}_{\mu}-\rho_{\mu})+\mu_{1}\gamma^{\nu}F_{\mu\nu}(\rho)\right)_{ba}\overline{H}^{\,(Q)}_a\right\rangle+h.c.\right]\nonumber\\
\end{eqnarray}
for the light vector meson exchange. In the above effective Lagrangians, the superfield $H^{(Q)}_a$ is defined as a linear combination of the pseudoscalar charmed meson $D$ with $I(J^P)=1/2(0^-)$ and vector charmed meson $D^*$ with $I(J^P)=1/2(1^-)$. And, the superfield $T^{(Q)\mu}_a$ is expressed as a combination of the axial-vector charmed meson $D_1$ with $I(J^P)=1/2(1^+)$ and tensor charmed meson $D_2^{*}$ with $I(J^P)=1/2(2^+)$ in the heavy quark limit \cite{Wise:1992hn}, which are defined by \cite{Ding:2008gr}
\begin{eqnarray}
H^{(Q)}_a&=&\frac{1+{v}\!\!\!\slash}{2}\left(D^{*(Q)\mu}_a\gamma_{\mu}-D^{(Q)}_a\gamma_5\right),\nonumber\\
T^{(Q)\mu}_a&=&\frac{1+{v}\!\!\!\slash}{2}\left(D^{*(Q)\mu\nu}_{2a}\gamma_{\nu}-\sqrt{\frac{3}{2}}D^{(Q)}_{1a\nu}\gamma_5\left(g^{\mu\nu}-\frac{\gamma^{\nu}(\gamma^{\mu}-v^{\mu})}{3}\right)\right),\nonumber\\
\end{eqnarray}
respectively. Here, $v^{\mu}=(1, \bf{0})$ denotes the four velocity under the non-relativistic approximation, and their conjugate fields satisfy $\overline{H}^{(Q)}_a=\gamma_0H^{(Q)\dagger}_a\gamma_0$ and $\overline{T}^{(Q)\mu}_a=\gamma_0T^{(Q)\mu\dagger}_a\gamma_0$ \cite{Ding:2008gr}. In addition, the covariant derivative $D_{\mu}$, the axial current $\mathcal{A}_\mu$, the vector current ${\cal V}_{\mu}$, the pseudo-Goldstone field $\xi$, the vector meson field $\rho_{\mu}$, and the vector meson field strength tensor $F^{\mu\nu}(\rho)$ can be defined as $D_{\mu}=\partial_{\mu}+\mathcal{V}_{\mu}$, ${\mathcal A}_{\mu}=\left(\xi^{\dagger}\partial_{\mu}\xi-\xi\partial_{\mu}\xi^{\dagger}\right)/2$, ${\mathcal V}_{\mu}=\left(\xi^{\dagger}\partial_{\mu}\xi+\xi\partial_{\mu}\xi^{\dagger}\right)/2$, $\xi=\text{exp}(i\mathbb{P}/f_{\pi})$, $\rho_{\mu}=ig_V\mathbb{V}_{\mu}/\sqrt{2}$, and $F^{\mu\nu}(\rho)=\partial^{\mu}\rho^{\nu}-\partial^{\nu}\rho^{\mu}+\left[\rho^{\mu},\rho^{\nu}\right]$, respectively. Here, the matrices for pseudoscalar mesons $\mathbb{P}$ and vector mesons $\mathbb{V}_{\mu}$ are expressed as \cite{Wang:2020bjt,Wang:2021hql,Chen:2018pzd}
\begin{eqnarray}
\left.\begin{array}{l}
{\mathbb{P}} = {\left(\begin{array}{ccc}
       \frac{\pi^0}{\sqrt{2}}+\frac{\eta}{\sqrt{6}} &\pi^+ &K^+\\
       \pi^-       &-\frac{\pi^0}{\sqrt{2}}+\frac{\eta}{\sqrt{6}} &K^0\\
       K^-         &\bar K^0   &-\sqrt{\frac{2}{3}} \eta     \end{array}\right)},\\
{\mathbb{V}}_{\mu} = {\left(\begin{array}{ccc}
       \frac{\rho^0}{\sqrt{2}}+\frac{\omega}{\sqrt{2}} &\rho^+ &K^{*+}\\
       \rho^-       &-\frac{\rho^0}{\sqrt{2}}+\frac{\omega}{\sqrt{2}} &K^{*0}\\
       K^{*-}         &\bar K^{*0}   & \phi     \end{array}\right)}_{\mu},
\end{array}\right.
\end{eqnarray}
respectively. At the leading order of the pseudo-Goldstone field $\xi$, the axial current $\mathcal{A}_\mu$ and the vector current ${\cal V}_{\mu}$ read as \cite{Wang:2019aoc}
\begin{eqnarray}
\mathcal{A}_\mu=\frac{i}{f_\pi}\partial_\mu{\mathbb P}, \quad {\mathrm{and}}\quad \mathcal{V}_{\mu}=0,
\end{eqnarray}
respectively. By expanding the compact effective Lagrangians, we can obtain  the detailed effective Lagrangians to describe the coupling involving the charmed mesons in the $H/T$-doublet and the light scalar, pseudoscalar, or vector mesons, which can be found in Refs. \cite{Wang:2019nwt,Wang:2019aoc,Wang:2020dya}.

As the important input parameters, the information about the coupling constants is crucial when studying the existence possibilities of the hadronic molecular states. These coupling constants can be extracted from the experimental data or calculated by the theoretical model \cite{Casalbuoni:1996pg,Wang:2019nwt,Falk:1992cx,Isola:2003fh,Cleven:2016qbn,Dong:2019ofp,He:2019csk,Wang:2019aoc,Wang:2020lua,Wang:2021aql,Wang:2020dya}, and the corresponding signs for several coupling constants can be determined by the quark model \cite{Riska:2000gd}. For example, the coupling constant $g$ is extracted from the experimental width of the $D^{*+}\rightarrow D^{0}\pi^{+}$ process \cite{Isola:2003fh}, the coupling constant $\beta$ can be determined from the vector meson dominance mechanism \cite{Isola:2003fh}, and the coupling constant $k$ can be estimated with the quark model \cite{Riska:2000gd}. Here, we need to indicate that these coupling constants are widely used to discuss the strong interactions \cite{Wang:2019nwt,Wang:2019aoc,Wang:2020dya,Wang:2021aql,Cleven:2016qbn,Dong:2019ofp,He:2019csk,Wang:2020lua}. In addition, the masses of these involved hadrons are taken from the Particle Data Group \cite{Zyla:2020zbs}. In Table \ref{parameters}, we collect these values of the coupling constants and hadron masses.
\renewcommand\tabcolsep{0.13cm}
\renewcommand{\arraystretch}{1.50}
\begin{table}[!htbp]
\caption{A summary of the coupling constants and hadron masses. Units of the coupling constants $h^{\prime}=(h_1+h_2)/\Lambda_{\chi}$, $\lambda$, $\lambda^{\prime\prime}$, and $\mu_1$ are $\rm{GeV}^{-1}$, the coupling constant $f_{\pi}$ is given in unit of GeV, and the masses of the hadrons are given in unit of MeV.}\label{parameters}
\centering
\begin{tabular}{llll}
\toprule[1.0pt]
\toprule[1.0pt]
$g_{\sigma}=-0.76$     &$g^{\prime\prime}_{\sigma}=0.76$                     &$|h^{\prime}_{\sigma}|=0.35$                 &$g=0.59$     \\
$k=-0.59$ &  $|h^{\prime}|=0.55$            &$f_{\pi}=0.132$ &$\beta=-0.90$               \\
$\beta^{\prime\prime}=0.90$                    &$\lambda=-0.56$& $\lambda^{\prime\prime}=0.56$                    &$|\zeta_1|=0.20$      \\
$\mu_1=0$           &$g_V=5.83$   & $m_{\sigma}=600.00$       &$m_{\pi}=137.27$  \\
$m_{\eta}=547.86$       &$m_{\rho}=775.26$& $m_{\omega}=782.66$    &$m_D=1867.24$ \\
$m_{D^*}=2008.56$   &$m_{D_{1}}=2422.00$ &$m_{D_{2}^{*}}=2463.05$     \\
\bottomrule[1.0pt]
\bottomrule[1.0pt]
\end{tabular}
\end{table}

In addition, we need the normalization relations for these discussed charmed mesons to write out the concrete scattering amplitude quantitatively, which include
\begin{eqnarray}
\left.\begin{array}{ll}
\langle 0|D|c\bar{q}(0^-)\rangle=\sqrt{m_{D}},\quad&\langle 0|D^{*\mu}|c\bar{q}(1^-)\rangle=\epsilon^\mu\sqrt{m_{D^*}},\\
\langle 0|D_{1}^{\mu}|c\bar{q}(1^+)\rangle=\epsilon^\mu\sqrt{m_{D_{1}}},\quad&\langle 0|D_{2}^{*\mu\nu}|c\bar{q}(2^+)\rangle=\zeta^{\mu\nu}\sqrt{m_{D_{2}^*}}.\\
\end{array}\right.
\end{eqnarray}
Here, the polarization vector $\epsilon^\mu_{m}\,(m=0,\,\pm1)$ describing spin-1 field can be expressed as $\epsilon_{0}^{\mu}= \left(0,0,0,-1\right)$ and $\epsilon_{\pm}^{\mu}= \left(0,\,\pm1,\,i,\,0\right)/\sqrt{2}$ in the static limit, and the polarization tensor $\zeta^{\mu\nu}_{m}(m =0,\,\pm1,\,\pm2)$ describing spin-2 field can be constructed as $\zeta^{\mu\nu}_{m}=\sum_{m1,m2}\langle 1, m_1; 1, m_2|2, m\rangle\epsilon^{\mu}_{m_1}\epsilon^{\nu}_{m_2}$ \cite{Cheng:2010yd}.

\subsection{The effective potential in the momentum (coordinate) space}
For the effective potential in the momentum space $\mathcal{V}^{h_1h_2\to h_3h_4}_E(\bm{q})$, we have
\begin{eqnarray}\label{breit}
\mathcal{V}_E^{h_1h_2\to h_3h_4}(\bm{q})=-\frac{\mathcal{M}(h_1h_2\to h_3h_4)} {\sqrt{2m_{h_1}2m_{h_2}2m_{h_3}2m_{h_4}}}
\end{eqnarray}
in terms of the Breit approximation \cite{Breit:1929zz,Breit:1930zza} and the nonrelativistic normalization.

In order to solve the coupled channel Schr\"{o}dinger equation in the coordinate space, the effective potential in the coordinate space $\mathcal{V}^{h_1h_2\to h_3h_4}_E(\bm{r})$ should be given, which can be obtained by performing Fourier transformation for the effective potential in the momentum space $\mathcal{V}^{h_1h_2\to h_3h_4}_E(\bm{q})$. The general relation can be explicitly expressed as
\begin{eqnarray}
\mathcal{V}^{h_1h_2\to h_3h_4}_E(\bm{r}) =\int \frac{d^3\bm{q}}{(2\pi)^3}e^{i\bm{q}\cdot\bm{r}}\mathcal{V}^{h_1h_2\to h_3h_4}_E(\bm{q})\mathcal{F}^2(q^2,m_E^2).
\end{eqnarray}
In the above Fourier transformation, $q$ and $m_E$ are the four-momentum and mass of the exchanged light mesons, respectively. Generally speaking, the monopole type form factor $\mathcal{F}(q^2,m_E^2) = (\Lambda^2-m_E^2)/(\Lambda^2-q^2)$ is introduced in each interactive vertex phenomenologically \cite{Tornqvist:1993ng,Tornqvist:1993vu}, which describes the finite size effect of the discussed hadrons and compensates the off-shell effect of the exchanged light mesons. Here, $\Lambda$ is the cutoff of the OBE potential \cite{Chen:2016qju}. The more explicit forms of the relevant Fourier transformation can be referred to Ref. \cite{Wang:2020dya}.

\subsection{The flavor and spin-orbital wave functions}
For the doubly-charmed molecular tetraquark states composed of the charmed meson in $H$-doublet and the charmed meson in $T$-doublet, their isospin quantum numbers are either $I=0$ or $1$. When discussing the concrete effective potentials of these investigated doubly-charmed tetraquark systems within the OBE model, we take the same form for the flavor wave functions as that given in the former work \cite{Li:2012ss,Chen:2021vhg,Liu:2019stu}. In Table~\ref{flavorwave}, we summarize the flavor wave functions $|I, I_3\rangle$ for these discussed doubly-charmed tetraquark $HT$ systems.
\renewcommand\tabcolsep{0.50cm}
\renewcommand{\arraystretch}{1.80}
\begin{table}[!htbp]
\caption{Flavor wave functions $|I, I_3\rangle$ for these discussed doubly-charmed tetraquark $HT$ systems. Here, $H=(D,\,D^{*})$, $T=(D_{1},\,D_{2}^{*})$, $I$ and $I_3$ represent their isospin and the third component of these discussed doubly-charmed tetraquark $HT$ systems, respectively.}\label{flavorwave}
\begin{tabular}{l|c}
\toprule[1.0pt]
\toprule[1.0pt]
$|I, I_3\rangle$&Configurations\\
\midrule[1.0pt]
$|1, 1\rangle$ &$\dfrac{1}{\sqrt{2}}\left(|H^{+}T^{+}\rangle+|T^{+}H^{+}\rangle\right)$\\
$|1,0\rangle$&$\dfrac{1}{2}\left[\left(|H^{0}T^{+}\rangle+|H^{+}T^{0}\rangle\right)+\left(|T^{+}H^{0}\rangle+|T^{0}H^{+}\rangle\right)\right]$\\
$|1, -1\rangle$&$\dfrac{1}{\sqrt{2}}\left(|H^{0}T^{0}\rangle+|T^{0}H^{0}\rangle\right)$\\
$|0,0\rangle$&$\dfrac{1}{2}\left[\left(|H^{0}T^{+}\rangle-|H^{+}T^{0}\rangle\right)-\left(|T^{+}H^{0}\rangle-|T^{0}H^{+}\rangle\right)\right]$\\
\bottomrule[1.0pt]
\bottomrule[1.0pt]
\end{tabular}
\end{table}

And then, the general expressions for the spin-orbital wave functions $|{}^{2S+1}L_{J}\rangle$ are written as
\begin{eqnarray}
|{\mathcal A}_{0} {\mathcal A}_{1}\rangle &=&\sum_{m,m_L}C^{J,M}_{1m,Lm_L}\epsilon_{m}^\mu|Y_{L,m_L}\rangle,\\
|{\mathcal A}_{0} {\mathcal A}_{2}\rangle &=&\sum_{m,m_L}C^{J,M}_{2m,Lm_L}\zeta_{m}^{\mu\nu}|Y_{L,m_L}\rangle,\\
|{\mathcal A}_{1} {\mathcal A}_{1}\rangle &=&\sum_{m,m^{\prime},m_S,m_L}C^{S,m_S}_{1m,1m^{\prime}}C^{J,M}_{Sm_S,Lm_L}\epsilon_{m}^\mu\epsilon_{m^{\prime}}^\nu|Y_{L,m_L}\rangle,\\
|{\mathcal A}_{1} {\mathcal A}_{2}\rangle &=& \sum_{m,m^{\prime},m_S,m_L}C^{S,m_S}_{1m,2m^{\prime}}C^{J,M}_{Sm_S,Lm_L}\epsilon_{m}^\lambda\zeta_{m^{\prime}}^{\mu\nu}|Y_{L,m_L}\rangle.
\end{eqnarray}
In the above expressions, the notations ${\mathcal A}_0$, ${\mathcal A}_1$, and ${\mathcal A}_2$ denote the charmed mesons $D$, $D^{*}(D_{1})$, and $D_{2}^{*}$, respectively, while $C^{e,f}_{ab,cd}$ and $|Y_{L,m_L}\rangle$ stand for the Clebsch-Gordan coefficient and the spherical harmonics function, respectively.

As is known well, the deuteron is a loosely bound state composed of a proton and a neutron, which may be regarded as the well-established molecular state \cite{Tornqvist:1993ng,Tornqvist:1993vu,Wang:2019nwt,Chen:2017jjn}. The study of the deuteron shows that the $S$-$D$ wave mixing effect may play important role in forming the loosely bound states \cite{Wang:2019nwt}. Thus, we consider the $S$-$D$ wave mixing effect in our calculation, where the relevant channels are summarized in Table~\ref{spin-orbit wave functions}. Here, we use the notation $|^{2S+1}L_J\rangle$ to simply denote the total spin $S$, relative angular momentum $L$, and total angular momentum $J$ for the corresponding doubly-charmed tetraquark systems, and the detailed information of finding out the bound state solutions for these $S$-wave $HT$ systems can be found in Ref. \cite{Wang:2019nwt} when considering the $S$-$D$ wave mixing effect and the coupled channel effect.
\renewcommand\tabcolsep{0.19cm}
\renewcommand{\arraystretch}{1.50}
\begin{table}[!htpb]
\centering
\caption{The quantum numbers $J^{P}$ and possible channels $|^{2S+1}L_J\rangle$ for the $S$-wave $HT$ systems. Here, ``$...$" means that the $S$-wave components in the corresponding states do not exist.}\label{spin-orbit wave functions}
\begin{tabular}{c|cccc}\toprule[1pt]\toprule[1pt]
 $J^{P}$&$DD_{1}$&$DD_{2}^{*}$&$D^{*}D_{1}$&$D^{*}D_{2}^{*}$\\\midrule[1.0pt]
$0^{-}$&$...$&$...$&$|{}^1\mathbb{S}_{0}\rangle/|{}^5\mathbb{D}_{0}\rangle$&$...$\\
$1^{-}$ &$|{}^3\mathbb{S}_{1}\rangle/|{}^3\mathbb{D}_{1}\rangle$&$...$&$|{}^3\mathbb{S}_{1}\rangle/|{}^{3,5}\mathbb{D}_{1}\rangle$&$|{}^3\mathbb{S}_{1}\rangle/|{}^{3,5,7}\mathbb{D}_{1}\rangle$\\
$2^{-}$&$...$&$|{}^5\mathbb{S}_{2}\rangle/|{}^5\mathbb{D}_{2}\rangle$&$|{}^5\mathbb{S}_{2}\rangle/|{}^{1,3,5}\mathbb{D}_{2}\rangle$&$|{}^5\mathbb{S}_{2}\rangle/|{}^{3,5,7}\mathbb{D}_{2}\rangle$\\
$3^{-}$&$...$&$...$&$...$&$|{}^7\mathbb{S}_{3}\rangle/|{}^{3,5,7}\mathbb{D}_{3}\rangle$\\
\bottomrule[1pt]\bottomrule[1pt]
\end{tabular}
\end{table}

Through the above procedure, we can obtain the general expressions of the effective potentials in the coordinate space for these investigated doubly-charmed molecular tetraquark $HT$ systems, which are collected in Appendix~\ref{app01}.

\section{Bound state solutions for these discussed $S$-wave $HT$ systems}\label{sec3}

Based on the obtained effective potentials in the coordinate space for these discussed doubly-charmed tetraquark systems, we solve the coupled channel Schr$\rm{\ddot{o}}$dinger equation, where the obtained bound state solutions include the binding energy $E$, the root-mean-square radius $r_{\rm RMS}$, and the probabilities for different components $P_i$, which may provide us the critical information to judge whether these discussed doubly-charmed molecular tetraquark states exist or not.

Before producing the numerical calculation, we want to specify three points:
\begin{enumerate}
  \item Since there exists the repulsive centrifugal potential $\ell(\ell+1)/2\mu r^2$, a hadronic state with the higher partial wave $\ell \geqslant 1$ is less likely to generate the bound state \cite{Chen:2016qju,Guo:2017jvc}. Thus, we  mainly focus on  the $S$-wave $HT$ systems in the current work.
  \item Only the cutoff is a free parameter in the OBE model \cite{Chen:2016qju}, and we attempt to find the loosely bound state solutions by changing the cutoff parameter in the range of $0.8$-$3.0~{\rm GeV}$ in the following numerical analysis \cite{Chen:2018pzd,Wang:2019aoc}. According to the experience of studying the deuteron, a loosely bound state with the cutoff parameter around 1.0 GeV can be suggested as the possible hadronic molecular candidate \cite{Tornqvist:1993ng,Tornqvist:1993vu,Wang:2019nwt,Chen:2017jjn}, which is widely regarded as a reasonable input parameter to study the hadronic molecular states \cite{Chen:2016qju,Liu:2019zoy}. Although the bound state properties of these investigated $S$-wave $HT$ systems depend on the cutoff parameter, we can predict possible doubly-charmed molecular tetraquark candidates at the qualitative level.
  \item In addition, when judging whether the loosely bound state is an ideal hadronic molecular candidate, we expect that the reasonable binding energy should be at most tens of MeV, and the typical root-mean-square radius should be larger than the size of all the included component hadrons \cite{Chen:2016qju,Chen:2017xat}. This is mainly because that the hadronic molecular state is a loosely bound state \cite{Chen:2016qju,Liu:2019zoy}.
\end{enumerate}

In the following, we analyze the bound state properties of these investigated $S$-wave $HT$ systems by performing the single channel, $S$-$D$ wave mixing, and coupled channel analysis.

\subsection{The $S$-wave $DD_{1}$ system}

For the $DD_{1}$ system, we notice that there does not exist the long-range $\pi$ exchange, but the $\sigma$, $\rho$, and $\omega$ exchanges are allowed within the OBE model because of the spin-parity conservation. In Table~\ref{sr1}, we present the bound state properties for the $S$-wave $DD_{1}$ system. In our numerical analysis, we firstly get the bound state solutions with the single channel analysis. Furthermore, we consider the contribution of the $S$-$D$ wave mixing effect and the coupled channel effect, and repeat the whole procedure of finding the bound state solutions of the $S$-wave $DD_{1}$ system.
\renewcommand\tabcolsep{0.32cm}
\renewcommand{\arraystretch}{1.50}
\begin{table}[!htbp]
\caption{Bound state solutions for the $S$-wave $DD_{1}$  system. The cutoff $\Lambda$, the binding energy $E$, and the root-mean-square radius $r_{RMS}$ are in units of $ \rm{GeV}$, $\rm {MeV}$, and $\rm {fm}$, respectively. Here, ``$\times$" indicates no bound state solutions when scanning the cutoff parameter 0.8-3.0 ${\rm GeV}$, and we label the major probability for the corresponding channels in a bold manner.}\label{sr1}
\begin{tabular}{c|cccc}\toprule[1.0pt]\toprule[1.0pt]
\multicolumn{5}{c}{Single channel analysis}\\\midrule[1.0pt]
$I(J^{P})$&$\Lambda$ &$E$&$r_{\rm RMS}$ \\
\multirow{3}{*}{$0(1^{-})$} &2.30&$-0.30$&4.88   \\
                            &2.65&$-1.98$&2.40 \\
                            &3.00&$-4.32$&1.70 \\
\multirow{1}{*}{$1(1^{-})$} &$\times$&$\times$&$\times$\\\midrule[1.0pt]
\multicolumn{5}{c}{$S$-$D$ wave mixing analysis}\\\midrule[1.0pt]
$I(J^{P})$&$\Lambda$&$E$&$r_{\rm RMS}$&$P({}^3\mathbb{S}_{1}/{}^3\mathbb{D}_{1})$  \\
\multirow{3}{*}{$0(1^{-})$} &2.25&$-0.29$&4.92&\textbf{99.96}/0.04    \\
                            &2.63&$-2.45$&2.19&\textbf{99.87}/0.13   \\
                            &3.00&$-5.56$&1.52&\textbf{99.78}/0.22    \\
\multirow{1}{*}{$1(1^{-})$} &$\times$&$\times$&$\times$&$\times$      \\\midrule[1.0pt]
\multicolumn{5}{c}{Coupled channel analysis}\\\midrule[1.0pt]
$I(J^{P})$&$\Lambda$&$E$&$r_{\rm RMS}$&$P(DD_{1}/D^{*}D_{1}/D^{*}D_{2}^{*})$  \\
\multirow{3}{*}{$0(1^{-})$}  &1.53&$-0.94$&2.89&\textbf{75.30}/$o(0)$/24.70\\
                             &1.54&$-4.12$&1.21&\textbf{47.56}/$o(0)$/\textbf{52.44}\\
                             &1.55&$-9.17$&0.73&30.26/$o(0)$/\textbf{69.74}\\
\multirow{1}{*}{$1(1^{-})$}  &$\times$&$\times$&$\times$&$\times$ \\
\bottomrule[1pt]\bottomrule[1pt]
\end{tabular}
\end{table}

For the $S$-wave $DD_{1}$ state with $I(J^{P})=1(1^{-})$, we cannot find the bound state solution when scanning the cutoff parameter $\Lambda=0.8$-$3.0~{\rm GeV}$, even if we consider the contribution from the $S$-$D$ wave mixing effect and the coupled channel effect. Therefore, there does not exist the $S$-wave $DD_{1}$ molecular state with $I(J^{P})=1(1^{-})$.

In contrast, we can obtain the bound state solution for the $S$-wave $DD_{1}$ state with $I(J^{P})=0(1^{-})$ by setting the cutoff parameter $\Lambda$ around 2.30 GeV when only considering the contribution of the $S$-wave channel, and the bound state solution can also be found when the cutoff parameter is fixed to be larger than 2.25 GeV after adding the contribution of the $D$-wave channel. Besides, we further include the coupled channel effect for the $S$-wave $DD_{1}$ coupled system with $I(J^{P})=0(1^{-})$. It is possible to find the bound state solution when we tune the cutoff parameter to be around 1.53 GeV, where the contribution of the $D^{*}D_{2}^{*}$ channel becomes obvious with increasing the cutoff value. Here, we notice that this state is the mixture of the $DD_{1}$ and $D^{*}D_{2}^{*}$ channels, where their probabilities are all over several tens of percents. Thus, our numerical results indicate that the contribution from the coupled channel effect may play an essential role in generating the $S$-wave $DD_{1}$ bound state with $I(J^{P})=0(1^{-})$. If the existence of the $S$-wave $DD_{1}$ molecular state with $I(J^{P})=0(1^{-})$ is possible, it can decay into the $DD$, $DD^*$, and $D^*D^*$ channels through the $P$-wave interaction. In addition, the three-body strong decay channels $DD\pi$, $DD^*\pi$, and $D^*D^*\pi$ are also allowed.

\subsection{The $S$-wave $DD_{2}^{*}$ system}

In this subsection, we present the cutoff parameter $\Lambda$, the binding energy $E$, the root-mean-square radius $r_{\rm RMS}$, and the probabilities of the individual channel $P_i$ for the $S$-wave $DD_{2}^{*}$ system in Table~\ref{sr2}.
\renewcommand\tabcolsep{0.31cm}
\renewcommand{\arraystretch}{1.50}
\begin{table}[!htbp]
\caption{Bound state solutions for the $S$-wave $DD_{2}^{*}$  system. Conventions are the same as Table~\ref{sr1}.}\label{sr2}
\begin{tabular}{c|cccc}\toprule[1.0pt]\toprule[1.0pt]
\multicolumn{5}{c}{Single channel analysis}\\\midrule[1.0pt]
$I(J^{P})$&$\Lambda$ &$E$&$r_{\rm RMS}$\\
\multirow{3}{*}{$0(2^{-})$} &1.66&$-0.36$&4.64 \\
                            &1.77&$-3.71$&1.80 \\
                            &1.88&$-12.61$&1.04\\
\multirow{1}{*}{$1(2^{-})$} &$\times$&$\times$&$\times$\\\midrule[1.0pt]
\multicolumn{5}{c}{$S$-$D$ wave mixing analysis}\\\midrule[1.0pt]
$I(J^{P})$&$\Lambda$      &$E$&$r_{\rm RMS}$&$P({}^5\mathbb{S}_{2}/{}^5\mathbb{D}_{2})$  \\
\multirow{3}{*}{$0(2^{-})$}         &1.64&$-0.30$&4.89&\textbf{99.98}/0.02  \\
                                    &1.75&$-3.72$&1.79&\textbf{99.95}/0.05  \\
                                    &1.85&$-12.12$&1.06&\textbf{99.89}/0.11 \\
\multirow{1}{*}{$1(2^{-})$}         &$\times$&$\times$&$\times$&$\times$    \\\midrule[1.0pt]
\multicolumn{5}{c}{Coupled channel analysis}\\\midrule[1.0pt]
$I(J^{P})$&$\Lambda$&$E$&$r_{\rm RMS}$&$P(DD_{2}^{*}/D^{*}D_{1}/D^{*}D_{2}^{*})$ \\
\multirow{3}{*}{$0(2^{-})$} &1.55&$-0.24$&5.08&\textbf{99.76}/0.24/$o(0)$\\
                           &1.65&$-3.84$&1.73&\textbf{98.84}/1.16/$o(0)$\\
                            &1.74&$-12.11$&1.02&\textbf{97.63}/2.37/$o(0)$\\
\multirow{1}{*}{$1(2^{-})$}&$\times$&$\times$&$\times$&$\times$ \\
\bottomrule[1pt]\bottomrule[1pt]
\end{tabular}
\end{table}

For the $S$-wave $DD_{2}^{*}$ state with $I(J^{P})=0(2^{-})$, there exist the loosely bound state solution when the cutoff parameter $\Lambda$ is fixed to be around 1.66 GeV or even larger value when performing the single channel analysis. If making comparison of the bound state solutions with and without the $S$-$D$ wave mixing effect, it is obvious that the $S$-$D$ wave mixing effect is not obvious when generating the $S$-wave $DD_{2}^{*}$ bound state with $I(J^{P})=0(2^{-})$, where the $D$-wave contribution is less than 1\%. After including the contribution of the coupled channel effect, we can obtain the bound state solution with the cutoff parameter $\Lambda$ above 1.55 GeV for the $S$-wave $DD_{2}^{*}$ coupled system with $I(J^{P})=0(2^{-})$, where the dominant channel is $DD_{2}^{*}$ with probability 97 percent and the remaining channels have very small probabilities. By including the $S$-$D$ mixing effect and the coupled channel effect step by step, the cutoff value becomes smaller when reproducing the same binding energy. Thus, the coupled channel effect plays positive role in the formation of the $S$-wave $DD_{2}^{*}$ bound state with $I(J^{P})=0(2^{-})$. If the cutoff parameter $\Lambda$ smaller than 1.6 GeV is a reasonable input, we may conclude that the $S$-wave $DD_{2}^{*}$ state with $I(J^{P})=0(2^{-})$ is the possible doubly-charmed molecular tetraquark candidate.

For the $S$-wave $DD_{2}^{*}$ state with $I(J^{P})=1(2^{-})$, there do not exist the bound state solution until we increase the cutoff parameter to be around 3.0 GeV, even if we take into account the contribution of the $S$-$D$ wave mixing effect and the coupled channel effect. Therefore, we conclude that our quantitative calculation does not support the existence of the $S$-wave $DD_{2}^{*}$ molecular state with $I(J^{P})=1(2^{-})$.

In the following, we also discuss the strong decay channels for the possible $S$-wave $DD_{2}^{*}$ molecular candidate with $I(J^{P})=0(2^{-})$. This state can decay into the $DD^*$ and $D^*D^*$ channels through the $P$-wave interaction, and the $DD_1$ decay channel can be suppressed since it is a typical $D$-wave decay process. Additionally, the three-body strong decay modes are the $DD\pi$, $DD^*\pi$, and $D^*D^*\pi$ channels.

\subsection{The $S$-wave $D^{*} D_{1}$ system}

Different from the former $S$-wave $DD_{1}$ and $DD_{2}^{*}$ states, the $S$-wave $D^{*} D_{1}$ states are more abundant due to the possibility of different quantum number combination. In Table \ref{sr3}, we collect the bound state properties for the $S$-wave $D^{*} D_{1}$ system. In particular, we consider the coupled channel effect for the $S$-wave $D^{*} D_{1}$ system in the present work, and find that the coupled channel effect provides slight contribution in forming the $S$-wave $D^{*} D_{1}$ bound states, which is similar to the $S$-wave $D^{*} \bar D_{1}$ system \cite{Wang:2021aql}  mainly determined by the non-diagonal matrix elements of the effective interactions \cite{Li:2012bt}.
\renewcommand\tabcolsep{0.06cm}
\renewcommand{\arraystretch}{1.50}
\begin{table}[!htbp]
\caption{Bound state solutions for the $S$-wave $D^{*} D_{1}$ system. Conventions are the same as Table~\ref{sr1}.}\label{sr3}
\begin{tabular}{c|ccc|cccc}\toprule[1pt]\toprule[1pt]
\multicolumn{1}{c|}{Effect}&\multicolumn{3}{c|}{Single channel}&\multicolumn{4}{c}{$S$-$D$ wave mixing effect}\\\midrule[1.0pt]
$I(J^{P})$&$\Lambda$ &$E$&$r_{\rm RMS}$&$\Lambda$ &$E$&$r_{\rm RMS}$&$P({}^1\mathbb{S}_{0}/{}^5\mathbb{D}_{0})$ \\
\multirow{3}{*}{$0(0^{-})$}        &0.96&$-0.36$&4.52      &0.94&$-0.24$&5.03&\textbf{99.59}/0.41\\
                                   &1.02&$-4.84$&1.55      &1.00&$-3.87$&1.74&\textbf{99.29}/0.71\\
                                   &1.07&$-13.19$&1.01      &1.06&$-12.90$&1.04&\textbf{99.32}/0.68\\
\multirow{1}{*}{$1(0^{-})$} &$\times$&$\times$&$\times$ &$\times$&$\times$&$\times$&$\times$ \\\midrule[1.0pt]
$I(J^{P})$&$\Lambda$ &$E$&$r_{\rm RMS}$&$\Lambda$ &$E$&$r_{\rm RMS}$&$P({}^3\mathbb{S}_{1}/{}^3\mathbb{D}_{1}/{}^5\mathbb{D}_{1})$ \\
\multirow{3}{*}{$0(1^{-})$}        &1.27&$-0.33$&4.55     &1.23&$-0.28$&4.79&\textbf{99.59}/0.40/0.01\\
                                   &1.35&$-4.23$&1.59     &1.32&$-4.46$&1.58&\textbf{99.22}/0.77/0.01\\
                                  &1.43&$-12.25$&1.00     &1.40&$-12.33$&1.02&\textbf{99.17}/0.82/0.01\\
\multirow{1}{*}{$1(1^{-})$} &$\times$&$\times$&$\times$ &$\times$&$\times$&$\times$&$\times$ \\\midrule[1.0pt]
$I(J^{P})$&$\Lambda$ &$E$&$r_{\rm RMS}$&$\Lambda$ &$E$&$r_{\rm RMS}$&$P({}^5\mathbb{S}_{2}/{}^1\mathbb{D}_{2}/{}^3\mathbb{D}_{2}/{}^5\mathbb{D}_{2})$ \\
\multirow{3}{*}{$0(2^{-})$}         &2.32&$-0.30$&4.99      &2.05&$-0.30$&5.00&\textbf{99.03}/0.11/$o(0)$/0.87\\
                                    &2.59&$-3.58$&1.95       &2.33&$-3.39$&2.03&\textbf{97.84}/0.24/$o(0)$/1.92\\
                                    &2.86&$-12.34$&1.18       &2.61&$-12.27$&1.21&\textbf{97.14}/0.41/$o(0)$/2.45\\
\multirow{1}{*}{$1(2^{-})$} &$\times$&$\times$&$\times$ &$\times$&$\times$&$\times$&$\times$ \\
\bottomrule[1pt]\bottomrule[1pt]
\end{tabular}
\end{table}

For the $S$-wave $D^{*} D_{1}$ states with $I(J^{P})=0(0^{-})$ and $0(1^{-})$, we can obtain the bound state solutions when choosing the cutoff parameters around 0.96 GeV and 1.27 GeV, respectively. Additionally, after considering the contribution of the $S$-$D$ wave mixing effect, the bound state solutions can also be found with the cutoff parameters around 0.94 GeV and 1.23 GeV, respectively. Since they have shallow binding energy and suitable root-mean-square radius under the reasonable range of the cutoff parameter, we conclude that the $S$-wave $D^{*} D_{1}$ states with $I(J^{P})=0(0^{-})$ and $0(1^{-})$ should be viewed as the ideal doubly-charmed molecular tetraquark candidates according to the experience of studying the deuteron \cite{Tornqvist:1993ng,Tornqvist:1993vu,Wang:2019nwt,Chen:2017jjn}, where the $\pi$ exchange provides strongly attractive force at the long-range for the $S$-wave $D^{*} D_{1}$ states with $I(J^{P})=0(0^{-},\,1^{-})$. For the $S$-wave $D^{*} D_{1}$ state with $I(J^{P})=0(2^{-})$, there exist weakly bound state solution when the cutoff parameter is taken around 2.30 GeV if we only consider the contribution of the $S$-wave channel. And then, we can also find weakly bound state solution with the cutoff parameter around 2.05 GeV when adding the contribution of the $D$-wave channels. However, such cutoff parameter is away from the reasonable range around 1.0 GeV \cite{Tornqvist:1993ng,Tornqvist:1993vu,Wang:2019nwt,Chen:2017jjn}. By comparing the bound state properties of the $S$-wave $D^{*} D_{1}$ molecular candidates with $I(J^{P})=0(0^{-},\,1^{-})$, it is obvious that the $S$-wave $D^{*} D_{1}$ state with $I(J^{P})=0(2^{-})$ as the doubly-charmed molecular tetraquark candidate is no priority. Based on the analysis mentioned above, it is clear that the $S$-wave isoscalar $D^{*} D_{1}$ states with the lower spin may be bound more tightly compared to those higher spin states, which is related to the obtained operator values $\mathcal{O}_k^{(\prime)}$ for the $S$-wave $D^{*} D_{1}$ states with the difference total angular momentum $J$ \cite{Chen:2015add}.

Contrary to the above situation,  when only considering the contribution of the $S$-wave channel, there do not exist the bound state solution with the cutoff parameter restricted to be below 3.0 GeV for the $S$-wave isovector $D^{*} D_{1}$ states. This situation almost keeps the same when the $S$-$D$ wave mixing effect is included. Thus, we exclude the possibilities of the existence of the $S$-wave isovector $D^{*} D_{1}$ molecular states. This can be easily understood since the $\pi$ exchange potential provides the repulsive force for the $S$-wave isovector $D^{*} D_{1}$ states with the lower spin.

To summarize, we can predict that the $S$-wave $D^{*} D_{1}$ states with $I(J^{P})=0(0^{-})$ and $0(1^{-})$ can be viewed as the ideal candidates of the doubly-charmed molecular tetraquark states for their reasonable cutoff value, binding energy, and root-mean-square radius \cite{Tornqvist:1993ng,Tornqvist:1993vu,Wang:2019nwt,Chen:2017jjn}. In the following, we also discuss their possible strong decay behaviors. For the $S$-wave $D^{*} D_{1}$ molecular candidates with $I(J^{P})=0(0^{-})$ and $0(1^{-})$, the allowed two-body and three-body strong decay channels include
\begin{eqnarray*}
\left.\begin{array}{ll}
0(0^{-}):&DD^*,\,D^*D^*,\,DD_2^*,\,DD\pi,\,DD^*\pi,\,D^*D^*\pi,\\
0(1^{-}):&DD,\,DD^*,\,D^*D^*,\,DD_1,\,DD_2^*,\,DD\pi,\,DD^*\pi,\,D^*D^*\pi,
\end{array}\right.
\end{eqnarray*}
respectively. Thus, we suggest the experiments should firstly carry out the search for these suggested doubly-charmed molecular tetraquark candidates in the future.

\subsection{The $S$-wave $D^{*}D_{2}^{*}$ system}

For the $S$-wave $D^{*}D_{2}^{*}$ system, the corresponding bound state properties are collected in Table~\ref{sr4} for all possible spin-parity configurations. In order to check the specific roles of the $S$-$D$ wave mixing effect, we present the numerical results without and with considering the $S$-$D$ wave mixing effect.
\renewcommand\tabcolsep{0.06cm}
\renewcommand{\arraystretch}{1.50}
\begin{table}[!htbp]
\caption{Bound state solutions for the $S$-wave $D^{*}D_{2}^{*}$ system. Conventions are the same as Table~\ref{sr1}.}\label{sr4}
\begin{tabular}{c|ccc|cccc}\toprule[1pt]\toprule[1pt]
\multicolumn{1}{c|}{Effect}&\multicolumn{3}{c|}{Single channel}&\multicolumn{4}{c}{$S$-$D$ wave mixing effect}\\\midrule[1.0pt]
$I(J^{P})$&$\Lambda$ &$E$&$r_{\rm RMS}$&$\Lambda$ &$E$&$r_{\rm RMS}$&$P({}^3\mathbb{S}_{1}/{}^3\mathbb{D}_{1}/{}^5\mathbb{D}_{1}/{}^7\mathbb{D}_{1})$ \\
\multirow{3}{*}{$0(1^{-})$}        &1.06&$-0.39$&4.34     &1.04&$-0.28$&4.75&\textbf{99.72}/0.17/$o(0)$/0.11\\
                                   &1.12&$-4.08$&1.63     &1.11&$-4.55$&1.57&\textbf{99.42}/0.42/$o(0)$/0.16\\
                                   &1.18&$-11.69$&1.04     &1.17&$-12.45$&1.03&\textbf{99.33}/0.54/$o(0)$/0.14\\
\multirow{1}{*}{$1(1^{-})$} &$\times$&$\times$&$\times$ &$\times$&$\times$&$\times$&$\times$ \\\midrule[1.0pt]
$I(J^{P})$&$\Lambda$ &$E$&$r_{\rm RMS}$&$\Lambda$ &$E$&$r_{\rm RMS}$&$P({}^5\mathbb{S}_{2}/{}^3\mathbb{D}_{2}/{}^5\mathbb{D}_{2}/{}^7\mathbb{D}_{2})$ \\
\multirow{3}{*}{$0(2^{-})$}         &1.60&$-0.32$&4.65     &1.58&$-0.37$&4.49&\textbf{99.77}/$o(0)$/0.23/$o(0)$\\
                                   &1.80&$-4.67$&1.55     &1.78&$-4.67$&1.56&\textbf{99.65}/$o(0)$/0.35/$o(0)$\\
                                   &1.99&$-12.28$&1.02     &1.97&$-12.20$&1.03&\textbf{99.69}/$o(0)$/0.31/$o(0)$\\
\multirow{1}{*}{$1(2^{-})$} &$\times$&$\times$&$\times$ &$\times$&$\times$&$\times$&$\times$ \\\midrule[1.0pt]
$I(J^{P})$&$\Lambda$ &$E$&$r_{\rm RMS}$&$\Lambda$ &$E$&$r_{\rm RMS}$&$P({}^7\mathbb{S}_{3}/{}^3\mathbb{D}_{3}/{}^5\mathbb{D}_{3}/{}^7\mathbb{D}_{3})$ \\
\multirow{3}{*}{$0(3^{-})$}         &$\times$&$\times$&$\times$     &2.66&$-0.29$&5.08&\textbf{99.16}/0.02/$o(0)$/0.02\\
                                    &$\times$&$\times$&$\times$     &2.83&$-0.86$&3.59&\textbf{98.78}/0.10/$o(0)$/1.12\\
                                    &$\times$&$\times$&$\times$     &3.00&$-1.78$&2.65&\textbf{98.46}/0.13/$o(0)$/1.41\\
\multirow{3}{*}{$1(3^{-})$}         &1.98&$-0.12$&5.26     &1.98&$-0.49$&3.70&\textbf{99.96}/0.02/$o(0)$/0.02\\
                                    &2.00&$-3.33$&1.48     &2.00&$-4.31$&1.30&\textbf{99.95}/0.03/$o(0)$/0.02\\
                                    &2.02&$-9.05$&0.90     &2.02&$-10.44$&0.84&\textbf{99.95}/0.04/$o(0)$/0.01\\
\bottomrule[1pt]\bottomrule[1pt]
\end{tabular}
\end{table}

From analyzing the bound state solutions for the $S$-wave $D^{*}D_{2}^{*}$ system presented in Table~\ref{sr4}, we summarize the following points:
\begin{itemize}
  \item For the $S$-wave $D^{*}D_{2}^{*}$ state with $I(J^{P})=0(1^{-})$, we can obtain the binding energies around a few MeV and the root-mean-square radii around a few fm with the cutoff parameter $\Lambda$ larger than 1.06 GeV, when only considering the contribution of the $S$-wave channel. Besides,  we can also get the bound state solution when the cutoff parameter $\Lambda$ is tuned larger than 1.04 GeV after considering the $S$-$D$ wave mixing effect. However, the contribution of the $D$-wave channels is very small. Borrowing the experience of studying deuteron \cite{Tornqvist:1993ng,Tornqvist:1993vu,Wang:2019nwt,Chen:2017jjn}, we may regard the $S$-wave $D^{*}D_{2}^{*}$ state with $I(J^{P})=0(1^{-})$ as the possible doubly-charmed molecular tetraquark candidate. For the $S$-wave $D^{*}D_{2}^{*}$ state with $I(J^{P})=1(1^{-})$, we fail to find the bound state solution with the cutoff parameter less than $3.0~{\rm GeV}$, even if the contribution of the $S$-$D$ wave mixing effect is introduced. Thus, the $S$-wave $D^{*}D_{2}^{*}$ state with $I(J^{P})=1(1^{-})$ cannot be bound together to form the hadronic molecular state.
  
  \item For the $S$-wave $D^{*}D_{2}^{*}$ state with $I(J^{P})=0(2^{-})$, when the cutoff parameter $\Lambda$ is slightly bigger than 1.60 GeV, we can obtain the bound state solution with shallow binding energy and suitable root-mean-square radius by performing single channel analysis. When we further consider the $S$-$D$ wave mixing effect, the loosely bound state solution can also be obtained if we tune the cutoff parameter $\Lambda$ larger than 1.58 GeV. Thus, the $S$-wave $D^{*}D_{2}^{*}$ state with $I(J^{P})=0(2^{-})$ seems to be the possible doubly-charmed molecular tetraquark candidate. For the $S$-wave $D^{*}D_{2}^{*}$ state with $I(J^{P})=1(2^{-})$, we cannot find the bound state solution corresponding to the cutoff parameter $0.8<\Lambda<3.0$ GeV, even if we consider the contribution of the $S$-$D$ wave mixing effect. Therefore, we can exclude the possibilities of the $S$-wave $D^{*}D_{2}^{*}$ state with $I(J^{P})=1(2^{-})$ as the doubly-charmed molecular tetraquark.
  
  \item For the $S$-wave $D^{*}D_{2}^{*}$ state with $I(J^{P})=0(3^{-})$, if the contribution of the $D$-wave channels does not considered, we should mention that the interactions are not sufficient to form the bound state until we increase the cutoff parameter $\Lambda$ to be around 3.0$~{\rm GeV}$. For the $S$-wave $D^{*}D_{2}^{*}$ state with $I(J^{P})=1(3^{-})$, we find that the binding energy can reach up to several MeV when taking the cutoff parameter $\Lambda$ to be around 2.00 GeV with the single channel analysis. However, when we further consider the contribution of the $S$-$D$ wave mixing effect, we can find that there exist the loosely bound state solutions for the $S$-wave $D^{*}D_{2}^{*}$ states with $I(J^{P})=0(3^{-})$ and $1(3^{-})$ if setting the cutoff parameters $\Lambda$ larger than 2.66 GeV and 1.98 GeV, respectively. Comparing the numerical results with and without considering the $S$-$D$ wave mixing effect, we notice that the $S$-$D$ wave mixing effect is important in the formation of the $S$-wave $D^{*}D_{2}^{*}$ bound state with $I(J^{P})=0(3^{-})$. However, such cutoff parameters deviate from the reasonable range around 1.0 GeV \cite{Tornqvist:1993ng,Tornqvist:1993vu,Wang:2019nwt,Chen:2017jjn}. Thus, it seems that the $S$-wave $D^{*}D_{2}^{*}$ states with $I(J^{P})=0(3^{-})$ and $1(3^{-})$ as the doubly-charmed molecular tetraquark candidates are no priority.
\end{itemize}

Additionally, we also provide the allowed two-body and three-body strong decay channels for these predicted doubly-charmed molecular tetraquark candidates. For the most promising $S$-wave $D^{*} D_{2}^{*}$ molecular candidate with $I(J^{P})=0(1^{-})$, the allowed two-body decay modes include the $DD$, $DD^*$, $D^*D^*$, $DD_{1}$, $DD_2^*,$ and $D^{*} D_{1}$ channels, and the possible three-body decay modes are the $DD\pi$, $DD^*\pi,$ and $D^*D^*\pi$ channels.  For the possible $S$-wave $D^{*} D_{2}^{*}$ molecular candidate with $I(J^{P})=0(2^{-})$, the allowed two-body and three-body decay modes are the $DD^*$, $D^*D^*$, $DD_1$, $DD_2^*$, $D^{*} D_{1}$, $DD\pi$, $DD^*\pi,$ and $D^*D^*\pi$ channels.

\section{Summary}\label{sec4}

Exploring the exotic hadronic state is an interesting and important research topic of the hadron physics. Very recently, the LHCb Collaboration reported a new structure $T_{cc}^+$ when analyzing the $D^0D^0\pi^+$ invariant mass spectrum \cite{Tcc:talk}. This doubly-charmed tetraquark $T_{cc}^+$ can be assigned as the isoscalar $DD^*$ molecular state with $J^{P}=1^{+}$ \cite{Li:2012ss,Xu:2017tsr,Li:2021zbw,Chen:2021vhg}, which is due to the mass of this doubly-charmed tetraquark state $T_{cc}^+$ locates just below the $DD^*$ threshold. To some extent, this new observation makes the study of the interactions between two charmed mesons becomes an intriguing research issue.

\begin{figure}[!htbp]
\centering
\begin{tabular}{c}
\includegraphics[width=0.48\textwidth]{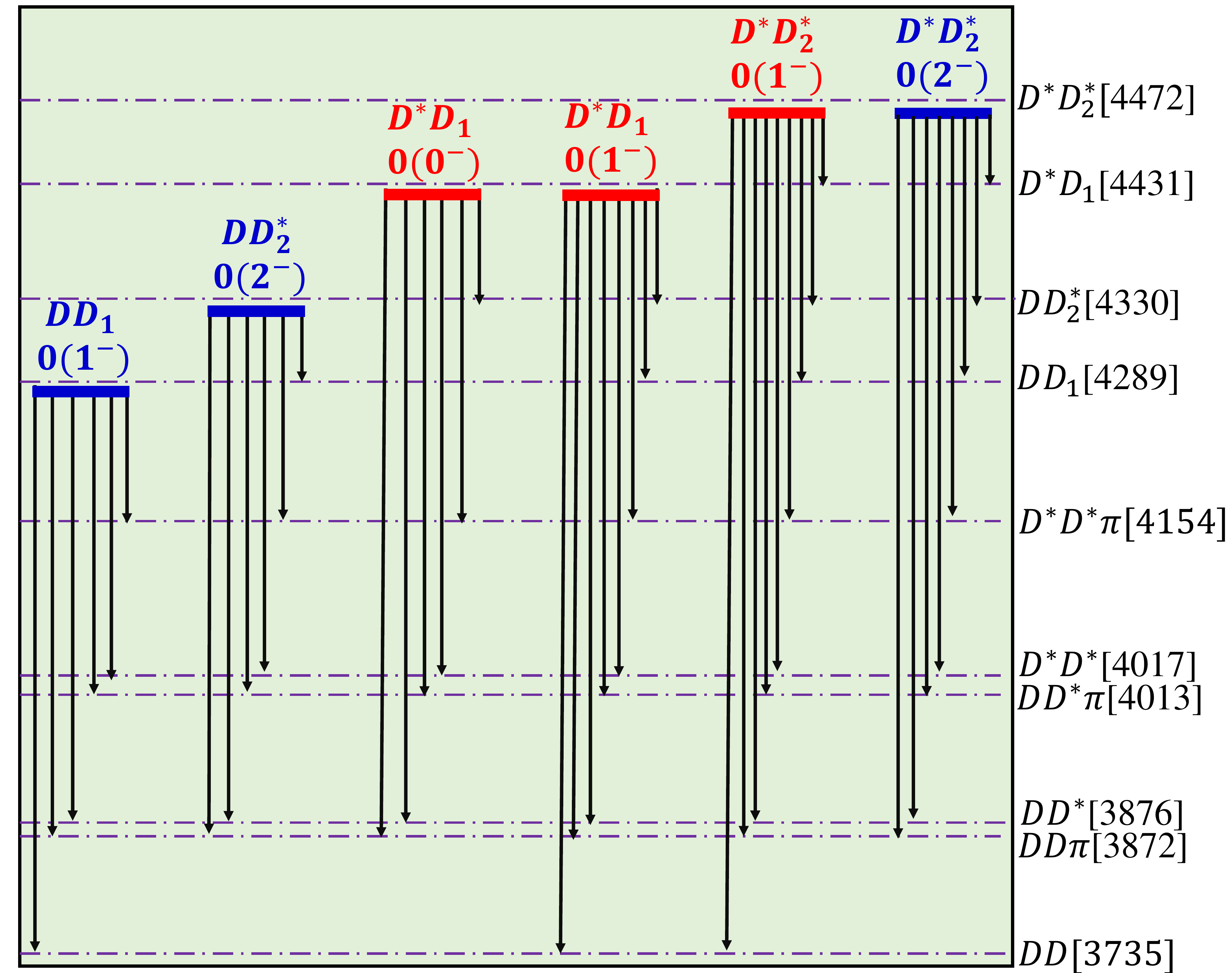}
\end{tabular}
\caption{Summary of two-body or three-body strong decay behaviors for these predicted new type of doubly-charmed molecular tetraquark states. Here, the dash-dotted lines denote the thresholds of the possible decay channels, and the red and blue thick solid lines represent these most promising and possible doubly-charmed molecular tetraquark candidates, respectively.}\label{Summary}
\end{figure}

In this work, we attempt to explore new type of the doubly-charmed molecular tetraquark candidates composed of the charmed meson in $H$-doublet and the charmed meson in $T$-doublet. In our concrete calculation, the effective potentials can be obtained by the OBE model when including the contribution from the $\sigma$, $\pi$, $\eta$, $\rho$, and $\omega$ exchanges, and we consider both the $S$-$D$ wave mixing effect and the coupled channel effect. Our numerical results show that the $S$-wave $D^{*} D_{1}$ states with $I(J^{P})=0(0^{-},\,1^{-})$ and the $S$-wave $D^{*}D_{2}^{*}$ state with $I(J^{P})=0(1^{-})$ should be viewed as the most promising doubly-charmed molecular tetraquark candidates, the $S$-wave $DD_{1}$ state with $I(J^{P})=0(1^{-})$, the $S$-wave $DD_{2}^{*}$ state with $I(J^{P})=0(2^{-})$, and the $S$-wave $D^{*}D_{2}^{*}$ state with $I(J^{P})=0(2^{-})$ are the possible candidates of the doubly-charmed molecular tetraquark states, and other investigated $S$-wave $HT$ states as the doubly-charmed molecular tetraquark candidates are no priority (see Fig. \ref{Summary}). Here, we need to emphasize that these possible doubly-charmed molecular tetraquark candidates have the typical exotic quark configuration $cc \bar q \bar q$ different from the conventional meson states, which gives us a good opportunity to identify the tetraquark hadronic states \cite{Wang:2021aql}.

Similarly to the observation of the $T_{cc}^+$ in the $D^0D^0\pi^+$ invariant mass spectrum \cite{Tcc:talk}, these predicted doubly-charmed molecular tetraquark states can be searched in these allowed two-body or three-body strong decay channels, since these predicted $S$-wave $HT$ molecular candidates can decay into the $DD$, $DD^*$, $D^*D^*$, $DD_1$, $DD_2^*,D^{*} D_{1}$, $DD\pi$, $DD^*\pi,$ and $D^*D^*\pi$ channels if the kinetically allowed (see Fig. \ref{Summary} for more details). These two-body or three-body strong decay modes can provide valuable information when searching for these predicted doubly-charmed molecular tetraquark states experimentally. As a potential experimental research issue, we strongly suggest the LHCb Collaboration to pay attention to these possible two-body or three-body strong decay channels from the proton-proton collisions with the accumulation of experimental data.  With more jigsaws of new types of the $T_{cc}$ states will be observed in future experiment,  the zoo of the doubly-charmed molecular tetraquark becomes booming.

\section*{ACKNOWLEDGMENTS}

This work is supported by the China National Funds for Distinguished Young Scientists under Grant No. 11825503, National Key Research and Development Program of China under Contract No. 2020YFA0406400, the 111 Project under Grant No. B20063, and the National Natural Science Foundation of China under Grant No. 12047501.

\appendix
\section{The effective potentials}\label{app01}

Before presenting the effective potentials for these investigated doubly-charmed tetraquark systems, we firstly define the function $Y(\Lambda_i,m_i,r)$, i.e.,
\begin{eqnarray}
Y_i\equiv \left\{
\begin{aligned}
|q_i|&\leqslant m,\ \frac{e^{-m_i r}-e^{-\Lambda^2_i r}}{4\pi r}-\frac{\Lambda^2_i-m^2_i}{8\pi\Lambda_i}e^{-\Lambda_i r};\\
|q_i|&>m,\ \frac{\mathrm{cos} (m^{\prime}_i r)-e^{-\Lambda_i r}}{4\pi r}-\frac{\Lambda^2_i+m^{\prime2}_i}{8\pi\Lambda_i}e^{-\Lambda_i r}.
\end{aligned}
\right.
\end{eqnarray}
Here, $m_i=\sqrt{m^2-q^2_i}$, $m^{\prime}_i=\sqrt{q^2_i-m^2}$, and $\Lambda_i=\sqrt{\Lambda^2-q^2_i}$. Additionally, the operators are defined as $\mathcal{Z}=\frac{1}{r^2}\frac{\partial}{\partial r}r^2\frac{\partial}{\partial r}$, $\mathcal{T}=r\frac{\partial}{\partial r}\frac{1}{r}\frac{\partial}{\partial r}$, and $\{\mathcal{T},\mathcal{Z}\}=\mathcal{T}\mathcal{Z}+\mathcal{Z}\mathcal{T}$.

For convenience, we define two functions $\mathcal{H}(I)Y(\Lambda,m_P,r)$ and $\mathcal{G}(I)Y(\Lambda,m_V,r)$ for these investigated doubly-charmed tetraquark systems, i.e.,
\begin{eqnarray}
&&\mathcal{H}(0)Y(\Lambda,m_P,r)=-\frac{3}{2}Y(\Lambda,m_{\pi},r)+\frac{1}{6}Y(\Lambda,m_{\eta},r),\\
&&\mathcal{H}(1)Y(\Lambda,m_P,r)=\frac{1}{2}Y(\Lambda,m_{\pi},r)+\frac{1}{6}Y(\Lambda,m_{\eta},r),\\
&&\mathcal{G}(0)Y(\Lambda,m_V,r)=-\frac{3}{2}Y(\Lambda,m_{\rho},r)+\frac{1}{2}Y(\Lambda,m_{\omega},r),\\
&&\mathcal{G}(1)Y(\Lambda,m_V,r)=\frac{1}{2}Y(\Lambda,m_{\rho},r)+\frac{1}{2}Y(\Lambda,m_{\omega},r).\label{vectorisopin}
\end{eqnarray}
Here, $\mathcal{H}(I)$ and $\mathcal{G}(I)$ are the isospin factors for these investigated doubly-charmed tetraquark systems, and $I$ denotes the isospin quantum number.

With the above preparation, the expressions of the effective potentials in the coordinate space for these investigated doubly-charmed tetraquark systems are given by
\begin{itemize}

\item For the process $D{D}_1\to D{D}_1$:
\begin{eqnarray}
\mathcal{V}&=&g_{\sigma}g^{\prime\prime}_{\sigma}\frac{\mathcal{O}_1+\mathcal{O}_1^{\prime}}{2}Y_{\sigma}-\frac{1}{2}\beta\beta^{\prime\prime}g^2_V\frac{\mathcal{O}_1+\mathcal{O}_1^{\prime}}{2}\mathcal{G}(I)Y_V.
\end{eqnarray}

\item For the process $D{D}_1\to{D}_1 D$:
\begin{eqnarray}
\mathcal{V}&=&\frac{2h^{\prime2}_{\sigma}}{9f^2_{\pi}}\left[\frac{\mathcal{O}_2+\mathcal{O}_2^{\prime}}{2}\mathcal{Z}+\frac{\mathcal{O}_3+\mathcal{O}_3^{\prime}}{2}\mathcal{T}\right]Y_{\sigma1}\nonumber\\
&&-\frac{\zeta^2_1g^2_V}{3}\frac{\mathcal{O}_2+\mathcal{O}_2^{\prime}}{2}\mathcal{G}(I)Y_{V1}.
\end{eqnarray}

\item For the process $D{D}^{\ast}_2\to D{D}^{\ast}_2$:
\begin{eqnarray}
\mathcal{V}&=&g_{\sigma}g^{\prime\prime}_{\sigma}\frac{\mathcal{O}_7+\mathcal{O}_7^{\prime}}{2}Y_{\sigma}-\frac{1}{2}\beta\beta^{\prime\prime}g^2_V\frac{\mathcal{O}_7+\mathcal{O}_7^{\prime}}{2}\mathcal{G}(I)Y_V.
\end{eqnarray}

\item For the process $D{D}^{\ast}_2\to{D}^{\ast}_2 D$:
\begin{eqnarray}
\mathcal{V}&=&-\frac{h^{\prime2}}{f^2_{\pi}}\left[\frac{\mathcal{O}_{8}+\mathcal{O}_{8}^{\prime}}{2}\mathcal{Z}\mathcal{Z}+\frac{\mathcal{O}_{9}+\mathcal{O}_{9}^{\prime}}{2}\mathcal{T}\mathcal{T}\right.\nonumber\\
&&\left.+\frac{\mathcal{O}_{10}+\mathcal{O}_{10}^{\prime}}{2}\{\mathcal{T},\mathcal{Z}\}\right]\mathcal{H}(I)Y_{P2}.
\end{eqnarray}

\item For the process $D^{\ast}{D}_1\to D^{\ast}{D}_1$:
\begin{eqnarray}
\mathcal{V}&=&g_{\sigma}g^{\prime\prime}_{\sigma}\frac{\mathcal{O}_{4}+\mathcal{O}_{4}^{\prime}}{2}Y_{\sigma}-\frac{1}{2}\beta\beta^{\prime\prime}g^2_V\frac{\mathcal{O}_{4}+\mathcal{O}_{4}^{\prime}}{2}\mathcal{G}(I)Y_V\nonumber\\
&&+\frac{5gk}{18f^2_{\pi}}\left[\frac{\mathcal{O}_{5}+\mathcal{O}_{5}^{\prime}}{2}\mathcal{Z}+\frac{\mathcal{O}_{6}+\mathcal{O}_{6}^{\prime}}{2}\mathcal{T}\right]\mathcal{H}(I)Y_P\nonumber\\
&&+\frac{5}{9}\lambda\lambda^{\prime\prime}g^2_V\left[2\frac{\mathcal{O}_{5}+\mathcal{O}_{5}^{\prime}}{2}\mathcal{Z}-\frac{\mathcal{O}_{6}+\mathcal{O}_{6}^{\prime}}{2}\mathcal{T}\right]\mathcal{G}(I)Y_V.\nonumber\\
\end{eqnarray}

\item For the process $D^{\ast}{D}_1\to{D}_1 D^{\ast}$:
\begin{eqnarray}
\mathcal{V}&=&-\frac{h^{\prime2}_{\sigma}}{18\pi^2_{\pi}}\left[\frac{\mathcal{O}_{5}+\mathcal{O}_{5}^{\prime}}{2}\mathcal{Z}+\frac{\mathcal{O}_{6}+\mathcal{O}_{6}^{\prime}}{2}\mathcal{T}\right]Y_{\sigma3}\nonumber\\
&&-\frac{h^{\prime2}}{6f^2_{\pi}}\left[\frac{\mathcal{O}_{11}+\mathcal{O}_{11}^{\prime}}{2}\mathcal{Z}\mathcal{Z}+\frac{\mathcal{O}_{12}+\mathcal{O}_{12}^{\prime}}{2}\mathcal{T}\mathcal{T}\right.\nonumber\\
&&\left.+\frac{\mathcal{O}_{13}+\mathcal{O}_{13}^{\prime}}{2}\{\mathcal{T},\mathcal{Z}\}\right]\mathcal{H}(I)Y_{P3}\nonumber\\
&&+\frac{\zeta^2_1g^2_V}{12}\frac{\mathcal{O}_{5}+\mathcal{O}_{5}^{\prime}}{2}\mathcal{G}(I)Y_{V3}.
\end{eqnarray}

\item For the process $D^{\ast}{D}^{\ast}_{2}\to D^{\ast}{D}^{\ast}_{2}$:
\begin{eqnarray}
\mathcal{V}&=&g_{\sigma}g^{\prime\prime}_{\sigma}\frac{\mathcal{O}_{14}+\mathcal{O}_{14}^{\prime}}{2}Y_{\sigma}-\frac{1}{2}\beta\beta^{\prime\prime}g^2_V\frac{\mathcal{O}_{14}+\mathcal{O}_{14}^{\prime}}{2}\mathcal{G}(I)Y_V\nonumber\\
&&+\frac{gk}{3f^2_{\pi}}\left[\frac{\mathcal{O}_{15}+\mathcal{O}_{15}^{\prime}}{2}\mathcal{Z}+\frac{\mathcal{O}_{16}+\mathcal{O}_{16}^{\prime}}{2}\mathcal{T}\right]\mathcal{H}(I)Y_P\nonumber\\
&&+\frac{2}{3}\lambda\lambda^{\prime\prime}g^2_V\left[2\frac{\mathcal{O}_{15}+\mathcal{O}_{15}^{\prime}}{2}\mathcal{Z}-\frac{\mathcal{O}_{16}+\mathcal{O}_{16}^{\prime}}{2}\mathcal{T}\right]\mathcal{G}(I)Y_V.\nonumber\\
\end{eqnarray}

\item For the process $D^{\ast}{D}^{\ast}_{2}\to{D}^{\ast}_{2}D^{\ast}$:
\begin{eqnarray}
\mathcal{V}&=&\frac{h^{\prime2}_{\sigma}}{3f^2_{\pi}}\left[\frac{\mathcal{O}_{17}+\mathcal{O}_{17}^{\prime}}{2}\mathcal{Z}+\frac{\mathcal{O}_{18}+\mathcal{O}_{18}^{\prime}}{2}\mathcal{T}\right]Y_{\sigma4}\nonumber\\
&&+\frac{h^{\prime2}}{f^2_{\pi}}\left[\frac{\mathcal{O}_{19}+\mathcal{O}_{19}^{\prime}}{2}\mathcal{Z}\mathcal{Z}+\frac{\mathcal{O}_{20}+\mathcal{O}_{20}^{\prime}}{2}\mathcal{T}\mathcal{T}\right.\nonumber\\
&&\left.+\frac{\mathcal{O}_{21}+\mathcal{O}_{21}^{\prime}}{2}\{\mathcal{T},\mathcal{Z}\}\right]\mathcal{H}(I)Y_{P4}\nonumber\\
&&-\frac{\zeta^2_1g^2_V}{2}\frac{\mathcal{O}_{17}+\mathcal{O}_{17}^{\prime}}{2}\mathcal{G}(I)Y_{V4}.
\end{eqnarray}
\end{itemize}
In the above expressions, the variables $q_i\,(i=1,\cdot\cdot\cdot,4)$ are written as $q_1=m_{D_1}-m_{D}$, $q_2=m_{D^{\ast}_2}-m_{D}$, $q_3=m_{D_1}-m_{D^{\ast}}$, and $q_4=m_{D^{\ast}_2}-m_{D^{\ast}}$.

In the above OBE effective potentials, we also introduce several operators $\mathcal{O}_k^{(\prime)}$ involved in this work \cite{Wang:2020dya,Wang:2021aql}, i.e.,
\begin{eqnarray}\label{op}
\mathcal{O}_{1}&=&{\bm\epsilon^{\dagger}_4}\cdot{\bm\epsilon_2},~~~~~~\mathcal{O}_{2}={\bm\epsilon^{\dagger}_3}\cdot{\bm\epsilon_2},~~~~~~\mathcal{O}_{3}=T({\bm\epsilon^{\dagger}_3},{\bm\epsilon_2}),\nonumber\\
\mathcal{O}_{4}&=&\left({\bm\epsilon^{\dagger}_3}\cdot{\bm\epsilon_1}\right)\left({\bm\epsilon^{\dagger}_4}\cdot{\bm\epsilon_2}\right),~~~~~~\mathcal{O}_{5}=\left({\bm\epsilon^{\dagger}_3}\times{\bm\epsilon_1}\right)\cdot\left({\bm\epsilon^{\dagger}_4}\times{\bm\epsilon_2}\right),\nonumber\\
\mathcal{O}_{6}&=&T({\bm\epsilon^{\dagger}_3}\times{\bm\epsilon_1},{\bm\epsilon^{\dagger}_4}\times{\bm\epsilon_2}),\nonumber\\
\mathcal{O}_{7}&=&\mathcal{\sum}\left({\bm\epsilon^{\dagger}_{4m}}\cdot{\bm\epsilon_{2a}}\right)\left({\bm\epsilon^{\dagger}_{4n}}\cdot {\bm\epsilon_{2b}}\right),\nonumber\\
\mathcal{O}_{8}&=&\frac{2}{27}\mathcal{\sum}\left({\bm\epsilon^{\dagger}_{3m}}\cdot{\bm\epsilon_{2a}}\right)\left({\bm\epsilon^{\dagger}_{3n}}\cdot {\bm\epsilon_{2b}}\right),\nonumber\\
\mathcal{O}_{9}&=&\frac{1}{27}\mathcal{\sum}T({\bm\epsilon^{\dagger}_{3m}},{\bm\epsilon^{\dagger}_{3n}})T({\bm\epsilon_{2a}},{\bm\epsilon_{2b}})\nonumber\\
&&+\frac{2}{27}\mathcal{\sum}T({\bm\epsilon^{\dagger}_{3m}},{\bm\epsilon_{2a}})T({\bm\epsilon^{\dagger}_{3n}},{\bm\epsilon_{2b}}),\nonumber\\
\mathcal{O}_{10}&=&\frac{2}{27}\mathcal{\sum}\left({\bm\epsilon^{\dagger}_{3m}}\cdot{\bm\epsilon_{2a}}\right)T({\bm\epsilon^{\dagger}_{3n}},{\bm\epsilon_{2b}}),\nonumber\\
\mathcal{O}_{11}&=&-\frac{1}{3}\left({\bm\epsilon^{\dagger}_3}\cdot{\bm\epsilon_1}\right)\left({\bm\epsilon^{\dagger}_4}\cdot{\bm\epsilon_2}\right)
+\frac{1}{3}\left({\bm\epsilon^{\dagger}_3}\cdot{\bm\epsilon^{\dagger}_4}\right)\left({\bm\epsilon_1}\cdot{\bm\epsilon_2}\right),\nonumber\\
\mathcal{O}_{12}&=&\frac{2}{3}T({\bm\epsilon^{\dagger}_3},{\bm\epsilon_1})T({\bm\epsilon^{\dagger}_4},{\bm\epsilon_2})+\frac{1}{3}T({\bm\epsilon^{\dagger}_3},{\bm\epsilon^{\dagger}_4})T({\bm\epsilon_1},{\bm\epsilon_2}),\nonumber\\
\mathcal{O}_{13}&=&\frac{1}{6}\left({\bm\epsilon^{\dagger}_3}\cdot{\bm\epsilon^{\dagger}_4}\right)T({\bm\epsilon_1},{\bm\epsilon_2})+\frac{1}{6}\left({\bm\epsilon_1}\cdot{\bm\epsilon_2}\right)T({\bm\epsilon^{\dagger}_3},{\bm\epsilon^{\dagger}_4})\nonumber\\
&&-\frac{1}{3}\left({\bm\epsilon^{\dagger}_3}\cdot{\bm\epsilon_1}\right)T({\bm\epsilon^{\dagger}_4},{\bm\epsilon_2}),\nonumber\\
\mathcal{O}_{14}&=&\mathcal{\sum}\left({\bm\epsilon^{\dagger}_3}\cdot{\bm\epsilon_1}\right)\left({\bm\epsilon^{\dagger}_{4m}}\cdot{\bm\epsilon_{2a}}\right)\left({\bm\epsilon^{\dagger}_{4n}}\cdot {\bm\epsilon_{2b}}\right),\nonumber\\
\mathcal{O}_{15}&=&\mathcal{\sum}\left({\bm\epsilon^{\dagger}_{4m}}\cdot{\bm\epsilon_{2a}}\right)\left[\left({\bm\epsilon^{\dagger}_{3}}\times{\bm\epsilon_{1}}\right)\cdot\left({\bm\epsilon^{\dagger}_{4n}}\times {\bm\epsilon_{2b}}\right)\right],\nonumber\\
\mathcal{O}_{16}&=&\mathcal{\sum}\left({\bm\epsilon^{\dagger}_{4m}}\cdot{\bm\epsilon_{2a}}\right)T({\bm\epsilon^{\dagger}_{3}}\times{\bm\epsilon_{1}},{\bm\epsilon^{\dagger}_{4n}}\times {\bm\epsilon_{2b}}),\nonumber\\
\mathcal{O}_{17}&=&\mathcal{\sum}\left({\bm\epsilon^{\dagger}_{3m}}\cdot{\bm\epsilon_1}\right)\left({\bm\epsilon^{\dagger}_4}\cdot{\bm\epsilon_{2a}}\right)\left({\bm\epsilon^{\dagger}_{3n}}\cdot {\bm\epsilon_{2b}}\right),\nonumber\\
\mathcal{O}_{18}&=&\mathcal{\sum}\left({\bm\epsilon^{\dagger}_{3m}}\cdot{\bm\epsilon_1}\right)\left({\bm\epsilon^{\dagger}_4}\cdot{\bm\epsilon_{2a}}\right)T({\bm\epsilon^{\dagger}_{3n}}, {\bm\epsilon_{2b}}),\nonumber\\
\mathcal{O}_{19}&=&\frac{1}{27}\mathcal{\sum}\left[\left({\bm\epsilon^{\dagger}_{3m}}\times{\bm\epsilon_{1}}\right)\cdot\left({\bm\epsilon^{\dagger}_{4}}\times{\bm\epsilon_{2a}}\right)\right]\left({\bm\epsilon^{\dagger}_{3n}}\cdot{\bm\epsilon_{2b}}\right)\nonumber\\
&&+\frac{1}{27}\mathcal{\sum}\left[\left({\bm\epsilon^{\dagger}_{3m}}\times{\bm\epsilon_{1}}\right)\cdot{\bm\epsilon_{2b}}\right]\left[{\bm\epsilon^{\dagger}_{3n}}\cdot\left({\bm\epsilon^{\dagger}_{4}}\times{\bm\epsilon_{2a}}\right)\right],\nonumber\\
\mathcal{O}_{20}&=&\frac{1}{27}\mathcal{\sum}T({\bm\epsilon^{\dagger}_{3m}}\times{\bm\epsilon_{1}},{\bm\epsilon^{\dagger}_{3n}})T({\bm\epsilon^{\dagger}_{4}}\times{\bm\epsilon_{2a}},{\bm\epsilon_{2b}})\nonumber\\
&&+\frac{1}{27}\mathcal{\sum}T({\bm\epsilon^{\dagger}_{3m}}\times{\bm\epsilon_{1}},{\bm\epsilon^{\dagger}_{4}}\times{\bm\epsilon_{2a}})T({\bm\epsilon^{\dagger}_{3n}},{\bm\epsilon_{2b}})\nonumber\\
&&+\frac{1}{27}\mathcal{\sum}T({\bm\epsilon^{\dagger}_{3m}}\times{\bm\epsilon_{1}},{\bm\epsilon_{2b}})T({\bm\epsilon^{\dagger}_{3n}},{\bm\epsilon^{\dagger}_{4}}\times{\bm\epsilon_{2a}}),\nonumber\\
\mathcal{O}_{21}&=&\frac{1}{54}\mathcal{\sum}\left[\left({\bm\epsilon^{\dagger}_{3m}}\times{\bm\epsilon_{1}}\right)\cdot\left({\bm\epsilon^{\dagger}_{4}}\times{\bm\epsilon_{2a}}\right)\right]T({\bm\epsilon^{\dagger}_{3n}},{\bm\epsilon_{2b}})\nonumber\\
&&+\frac{1}{54}\mathcal{\sum}\left[\left({\bm\epsilon^{\dagger}_{3m}}\times{\bm\epsilon_{1}}\right)\cdot{\bm\epsilon_{2b}}\right]T({\bm\epsilon^{\dagger}_{3n}},{\bm\epsilon^{\dagger}_{4}}\times{\bm\epsilon_{2a}})\nonumber\\
&&+\frac{1}{54}\mathcal{\sum}\left({\bm\epsilon^{\dagger}_{3n}}\cdot{\bm\epsilon_{2b}}\right)T({\bm\epsilon^{\dagger}_{3m}}\times{\bm\epsilon_{1}},{\bm\epsilon^{\dagger}_{4}}\times{\bm\epsilon_{2a}})\nonumber\\
&&+\frac{1}{54}\mathcal{\sum}\left[{\bm\epsilon^{\dagger}_{3n}}\cdot\left({\bm\epsilon^{\dagger}_{4}}\times{\bm\epsilon_{2a}}\right)\right]T({\bm\epsilon^{\dagger}_{3m}}\times{\bm\epsilon_{1}},{\bm\epsilon_{2b}}).
\end{eqnarray}
Here, we define $\mathcal{\sum}=\sum_{m,n,a,b}C^{2,m+n}_{1m,1n}C^{2,a+b}_{1a,1b}$, and $T({\bm x},{\bm y})= 3\left(\hat{\bm r} \cdot {\bm x}\right)\left(\hat{\bm r} \cdot {\bm y}\right)-{\bm x} \cdot {\bm y}$ with $\hat{\bm r}={\bm r}/|{\bm r}|$ is the tensor force operator. For the operators $\mathcal{O}_k^{\prime}$, we just need to make the change with the subscripts of the polarization vector $\epsilon_i$ for the operators $\mathcal{O}_k$, i.e., $1 \leftrightarrow 2$ and $3 \leftrightarrow 4$. In our calculation, the corresponding matrices elements $\langle f|\mathcal{O}_k^{(\prime)}|i\rangle$ are obtained by sandwiched these operators $\mathcal{O}_k^{(\prime)}$ between the relevant spin-orbit wave functions of the initial and final states. In Tables~\ref{matrix1}-\ref{matrix2}, we collect the obtained operator matrix elements $\mathcal{O}_k^{(\prime)}[J]\,(k=1,\cdot\cdot\cdot,21)$, which will be used in the calculation \cite{Wang:2020dya,Wang:2021aql}.
\renewcommand\tabcolsep{0.01cm}
\renewcommand{\arraystretch}{1.50}
\begin{table*}[!htbp]
\caption{The relevant operator matrix elements $\mathcal{O}_k^{(\prime)}[J]=\langle f|\mathcal{O}_k^{(\prime)}|i\rangle\,(k=1,\cdot\cdot\cdot,13)$.}\label{matrix1}
\begin{tabular}{l|l|l}\toprule[1pt]\toprule[1pt]
\multicolumn{3}{c}{$\mathcal{O}_k^{(\prime)}[J]=\langle f|\mathcal{O}_k^{(\prime)}|i\rangle\,(k=1,\cdot\cdot\cdot,13)$}\\\midrule[1.0pt]
$\begin{array}{l}\mathcal{O}_1^{(\prime)}[1]=\rm {diag}(1,1)\\ \mathcal{O}_2^{(\prime)}[1]=\rm {diag}(1,1)\end{array}$$\mathcal{O}_{3}^{(\prime)}[1]=\left(\begin{array}{cc} 0 & -\sqrt{2} \\ -\sqrt{2} & 1\end{array}\right)$ &$\begin{array}{l}\mathcal{O}_4^{(\prime)}[0]=\rm {diag}(1,1)\\ \mathcal{O}_5^{(\prime)}[0]=\rm {diag}(2,-1) \end{array}$$\mathcal{O}_{6}^{(\prime)}[0]=\left(\begin{array}{cc} 0 & \sqrt{2} \\ \sqrt{2} & 2\end{array}\right)$ & $\begin{array}{l} \mathcal{O}_4^{(\prime)}[1]=\rm {diag}(1,1,1) \\ \mathcal{O}_5^{(\prime)}[1]=\rm {diag}(1,1,-1) \end{array}$\\
$\mathcal{O}_{6}^{(\prime)}[1]=\left(\begin{array}{ccc} 0 & -\sqrt{2} &0 \\ -\sqrt{2} & 1 &0 \\ 0&0&1\end{array}\right)$&$\begin{array}{l} \mathcal{O}_4^{(\prime)}[2]=\rm {diag}(1,1,1,1) \\ \mathcal{O}_5^{(\prime)}[2]=\rm {diag}(-1,2,1,-1) \\ \mathcal{O}_7^{(\prime)}[2]=\rm {diag}(1,1)\\ \mathcal{O}_8[2]=\rm {diag}(\frac{2}{27},\frac{2}{27})\end{array}$ & $\mathcal{O}_{6}^{(\prime)}[2]=\left(\begin{array}{cccc} 0 & \frac{\sqrt{2}}{\sqrt{5}} & 0 &-\frac{\sqrt{14}}{\sqrt{5}} \\ \frac{\sqrt{2}}{\sqrt{5}} & 0 & 0 &-\frac{2}{\sqrt{7}} \\ 0 & 0 &-1 &0 \\ -\frac{\sqrt{14}}{\sqrt{5}}&-\frac{2}{\sqrt{7}}&0&-\frac{3}{7}\end{array}\right)$ \\
$\mathcal{O}_9^{(\prime)}[2]=\left(\begin{array}{cc} \frac{8}{135} & -\frac{4\sqrt{2}}{27\sqrt{35}} \\ -\frac{4\sqrt{2}}{27\sqrt{35}} & \frac{4}{27}\end{array}\right)$ & $\mathcal{O}_{10}^{(\prime)}[2]=\left(\begin{array}{cc} 0 & -\frac{\sqrt{7}}{27\sqrt{10}} \\ -\frac{\sqrt{7}}{27\sqrt{10}} & -\frac{1}{126}\end{array}\right)$&$\begin{array}{l} \mathcal{O}_{11}^{(\prime)}[0]=\rm {diag}(\frac{2}{3},-\frac{1}{3})\\ \mathcal{O}_{11}^{(\prime)}[1]=\rm {diag}(-\frac{1}{3},-\frac{1}{3},-\frac{1}{3})\\\mathcal{O}_{11}^{(\prime)}[2]=\rm {diag}(-\frac{1}{3},\frac{2}{3},-\frac{1}{3},-\frac{1}{3})\end{array}$ \\
$\mathcal{O}_{12}^{(\prime)}[0]=\left(\begin{array}{cc} \frac{4}{3} & -\frac{2\sqrt{2}}{3} \\ -\frac{2\sqrt{2}}{3} & 4\end{array}\right)$&$\mathcal{O}_{13}^{(\prime)}[0]=\left(\begin{array}{cc} 0 & \frac{\sqrt{2}}{15} \\ -\frac{8\sqrt{2}}{15} & -\frac{1}{15}\end{array}\right)$&$\mathcal{O}_{12}^{(\prime)}[1]=\left(\begin{array}{ccc} -\frac{2}{3} & -\frac{2\sqrt{2}}{3} &0 \\ -\frac{2\sqrt{2}}{3} & 0 &0 \\ 0&0&-\frac{4}{3}\end{array}\right)$\\
$\mathcal{O}_{13}^{(\prime)}[1]=\left(\begin{array}{ccc} 0 & -\frac{1}{30\sqrt{2}}&\frac{\sqrt{3}}{10\sqrt{2}} \\ -\frac{1}{30\sqrt{2}}&\frac{4}{105} &\frac{3\sqrt{3}}{70} \\ \frac{\sqrt{3}}{10\sqrt{2}}&\frac{3\sqrt{3}}{70}&-\frac{1}{105}\end{array}\right)$&$\mathcal{O}_{12}^{(\prime)}[2]=\left(\begin{array}{cccc} \frac{8}{15} &-\frac{2\sqrt{2}}{3\sqrt{5}} & 0 &-\frac{4\sqrt{2}}{3\sqrt{35}} \\-\frac{2\sqrt{2}}{3\sqrt{5}} &\frac{4}{3} & 0 &\frac{4}{3\sqrt{7}} \\ 0 &0 & -\frac{4}{3} &0 \\ -\frac{4\sqrt{2}}{3\sqrt{35}}&\frac{4}{3\sqrt{7}}&0&\frac{4}{3}\end{array}\right)$&$\mathcal{O}_{13}^{(\prime)}[2]=\left(\begin{array}{cccc} 0 &-\frac{1}{\sqrt{10}}& 0&\frac{\sqrt{2}}{3\sqrt{35}} \\ 0 &0& 0&-\frac{4}{21\sqrt{7}} \\0&0&-\frac{1}{14} &\frac{1}{14\sqrt{7}} \\ \frac{\sqrt{2}}{3\sqrt{35}}&\frac{23}{21\sqrt{7}}&\frac{1}{14\sqrt{7}}&\frac{13}{294}\end{array}\right)$\\
\bottomrule[1pt]\bottomrule[1pt]
\end{tabular}
\end{table*}

\renewcommand\tabcolsep{0.01cm}
\renewcommand{\arraystretch}{1.50}
\begin{table*}[!htbp]
\caption{The relevant operator matrix elements $\mathcal{O}_k^{(\prime)}[J]=\langle f|\mathcal{O}_k^{(\prime)}|i\rangle\,(k=14,\cdot\cdot\cdot,21)$.}\label{matrix2}
\begin{tabular}{l|l|l}\toprule[1pt]\toprule[1pt]
\multicolumn{3}{c}{$\mathcal{O}_k^{(\prime)}[J]=\langle f|\mathcal{O}_k^{(\prime)}|i\rangle\,(k=14,\cdot\cdot\cdot,21)$}\\\midrule[1.0pt]
$\begin{array}{l} \mathcal{O}_{14}^{(\prime)}[1]=\rm {diag}(1,1,1,1) \\ \mathcal{O}_{15}^{(\prime)}[1]=\rm {diag}(\frac{3}{2},\frac{3}{2},\frac{1}{2},-1)\\\mathcal{O}_{17}^{(\prime)}[1]=\rm {diag}(\frac{1}{6},\frac{1}{6},\frac{1}{2},1)\\\mathcal{O}_{19}^{(\prime)}[1]=\rm {diag}(\frac{1}{18},\frac{1}{18},\frac{5}{54},-\frac{1}{27})\end{array}$& $\mathcal{O}_{16}[1]=\left(\begin{array}{cccc} 0&\frac{3}{5\sqrt{2}}&\frac{\sqrt{6}}{\sqrt{5}}&\frac{\sqrt{21}}{5\sqrt{2}} \\ \frac{3}{5\sqrt{2}}&-\frac{3}{10}&\frac{\sqrt{3}}{\sqrt{5}}&-\frac{\sqrt{3}}{5\sqrt{7}}\\\frac{\sqrt{6}}{\sqrt{5}}&\frac{\sqrt{3}}{\sqrt{5}}&\frac{1}{2}&\frac{2}{\sqrt{35}}\\\frac{\sqrt{21}}{5\sqrt{2}}
&-\frac{\sqrt{3}}{5\sqrt{7}}&\frac{2}{\sqrt{35}}&\frac{48}{35}\end{array}\right)$&
$\mathcal{O}_{16}^{\prime}[1]=\left(\begin{array}{cccc} 0&\frac{3}{5\sqrt{2}}&-\frac{\sqrt{6}}{\sqrt{5}}&\frac{\sqrt{21}}{5\sqrt{2}} \\ \frac{3}{5\sqrt{2}}&-\frac{3}{10}&-\frac{\sqrt{3}}{\sqrt{5}}&-\frac{\sqrt{3}}{5\sqrt{7}}\\-\frac{\sqrt{6}}{\sqrt{5}}&-\frac{\sqrt{3}}{\sqrt{5}}&\frac{1}{2}&-\frac{2}{\sqrt{35}}\\\frac{\sqrt{21}}{5\sqrt{2}}
&-\frac{\sqrt{3}}{5\sqrt{7}}&-\frac{2}{\sqrt{35}}&\frac{48}{35}\end{array}\right)$\\
 $\mathcal{O}_{18}[1]=\left(\begin{array}{cccc} 0&-\frac{23}{15\sqrt{2}}&-\frac{2\sqrt{2}}{\sqrt{15}}&-\frac{\sqrt{7}}{5\sqrt{6}} \\ -\frac{23}{15\sqrt{2}}&\frac{23}{30}&-\frac{2}{\sqrt{15}}&\frac{1}{5\sqrt{21}}\\\frac{2\sqrt{2}}{\sqrt{15}}&\frac{2}{\sqrt{15}}&\frac{1}{2}&-\frac{2}{\sqrt{35}}\\-\frac{\sqrt{7}}{5\sqrt{6}}
&\frac{1}{5\sqrt{21}}&\frac{2}{\sqrt{35}}&\frac{24}{35}\end{array}\right)$&$\mathcal{O}_{18}^{\prime}[1]=\left(\begin{array}{cccc} 0&-\frac{23}{15\sqrt{2}}&-\frac{2\sqrt{2}}{\sqrt{15}}&\frac{\sqrt{7}}{5\sqrt{6}} \\ -\frac{23}{15\sqrt{2}}&\frac{23}{30}&\frac{2}{\sqrt{15}}&\frac{1}{5\sqrt{21}}\\-\frac{2\sqrt{2}}{\sqrt{15}}&-\frac{2}{\sqrt{15}}&\frac{1}{2}&\frac{2}{\sqrt{35}}\\-\frac{\sqrt{7}}{5\sqrt{6}}
&\frac{1}{5\sqrt{21}}&-\frac{2}{\sqrt{35}}&\frac{24}{35}\end{array}\right)$& $\mathcal{O}_{20}[1]=\left(\begin{array}{cccc} \frac{2}{45}&-\frac{\sqrt{2}}{45}&0&\frac{\sqrt{2}}{15\sqrt{21}} \\ -\frac{\sqrt{2}}{45}&\frac{1}{15}&0&-\frac{8}{15\sqrt{21}}\\0&0&-\frac{1}{9}&-\frac{2\sqrt{5}}{27\sqrt{7}}\\\frac{\sqrt{2}}{15\sqrt{21}}
&-\frac{8}{15\sqrt{21}}&\frac{2\sqrt{5}}{27\sqrt{7}}&\frac{92}{945}\end{array}\right)$\\
$\mathcal{O}_{20}^{\prime}[1]=\left(\begin{array}{cccc} \frac{2}{45}&-\frac{\sqrt{2}}{45}&0&\frac{\sqrt{2}}{15\sqrt{21}} \\ -\frac{\sqrt{2}}{45}&\frac{1}{15}&0&-\frac{8}{15\sqrt{21}}\\0&0&-\frac{1}{9}&\frac{2\sqrt{5}}{27\sqrt{7}}\\\frac{\sqrt{2}}{15\sqrt{21}}
&-\frac{8}{15\sqrt{21}}&-\frac{2\sqrt{5}}{27\sqrt{7}}&\frac{92}{945}\end{array}\right)$& $\mathcal{O}_{21}[1]=\left(\begin{array}{cccc} 0&-\frac{7}{90\sqrt{2}}&0&\frac{\sqrt{7}}{30\sqrt{6}}\\ -\frac{7}{90\sqrt{2}}&\frac{7}{180}&0&-\frac{1}{30\sqrt{21}}\\0&0&\frac{1}{108}&\frac{\sqrt{5}}{27\sqrt{7}}\\\frac{\sqrt{7}}{30\sqrt{6}}&-\frac{1}{30\sqrt{21}}&-\frac{\sqrt{5}}{27\sqrt{7}}&\frac{4}{315}\end{array}\right)$&$\mathcal{O}_{21}^{\prime}[1]=\left(\begin{array}{cccc} 0&-\frac{7}{90\sqrt{2}}&0&\frac{\sqrt{7}}{30\sqrt{6}}\\ -\frac{7}{90\sqrt{2}}&\frac{7}{180}&0&-\frac{1}{30\sqrt{21}}\\0&0&\frac{1}{108}&-\frac{\sqrt{5}}{27\sqrt{7}}\\\frac{\sqrt{7}}{30\sqrt{6}}&-\frac{1}{30\sqrt{21}}&\frac{\sqrt{5}}{27\sqrt{7}}&\frac{4}{315}\end{array}\right)$\\
$\begin{array}{l} \mathcal{O}_{14}^{(\prime)}[2]=\rm {diag}(1,1,1,1) \\ \mathcal{O}_{15}^{(\prime)}[2]=\rm {diag}(\frac{1}{2},\frac{3}{2},\frac{1}{2},-1)\\ \mathcal{O}_{17}^{(\prime)}[2]=\rm {diag}(\frac{1}{2},\frac{1}{6},\frac{1}{2},1) \\\mathcal{O}_{19}^{(\prime)}[2]=\rm {diag}(\frac{5}{54},\frac{1}{18},\frac{5}{54},-\frac{1}{27})\end{array}$&$\mathcal{O}_{16}[2]=\left(\begin{array}{cccc} 0&-\frac{3\sqrt{2}}{5}&-\frac{\sqrt{7}}{\sqrt{10}}&\frac{\sqrt{7}}{5} \\ -\frac{3\sqrt{2}}{5}&\frac{3}{10}&\frac{3}{\sqrt{35}}&-\frac{3\sqrt{2}}{5\sqrt{7}}\\-\frac{\sqrt{7}}{\sqrt{10}}&\frac{3}{\sqrt{35}}&-\frac{3}{14}&\frac{4\sqrt{2}}{7\sqrt{5}}\\\frac{\sqrt{7}}{5}
&-\frac{3\sqrt{2}}{5\sqrt{7}}&\frac{4\sqrt{2}}{7\sqrt{5}}&\frac{12}{35}\end{array}\right)$&$\mathcal{O}_{16}^{\prime}[2]=\left(\begin{array}{cccc} 0&\frac{3\sqrt{2}}{5}&-\frac{\sqrt{7}}{\sqrt{10}}&-\frac{\sqrt{7}}{5} \\ \frac{3\sqrt{2}}{5}&\frac{3}{10}&-\frac{3}{\sqrt{35}}&-\frac{3\sqrt{2}}{5\sqrt{7}}\\-\frac{\sqrt{7}}{\sqrt{10}}&-\frac{3}{\sqrt{35}}&-\frac{3}{14}&-\frac{4\sqrt{2}}{7\sqrt{5}}\\-\frac{\sqrt{7}}{5}
&-\frac{3\sqrt{2}}{5\sqrt{7}}&-\frac{4\sqrt{2}}{7\sqrt{5}}&\frac{12}{35}\end{array}\right)$\\
$\mathcal{O}_{18}[2]=\left(\begin{array}{cccc} 0&-\frac{2\sqrt{2}}{5}&-\frac{\sqrt{7}}{\sqrt{10}}&-\frac{\sqrt{7}}{5} \\ \frac{2\sqrt{2}}{5}&-\frac{23}{30}&-\frac{2}{\sqrt{35}}&\frac{\sqrt{2}}{5\sqrt{7}}\\-\frac{\sqrt{7}}{\sqrt{10}}&\frac{2}{\sqrt{35}}&-\frac{3}{14}&\frac{4\sqrt{2}}{7\sqrt{5}}\\\frac{\sqrt{7}}{5}
&\frac{\sqrt{2}}{5\sqrt{7}}&\frac{4\sqrt{2}}{7\sqrt{5}}&\frac{6}{35}\end{array}\right)$&$\mathcal{O}_{18}^{\prime}[2]=\left(\begin{array}{cccc} 0&\frac{2\sqrt{2}}{5}&-\frac{\sqrt{7}}{\sqrt{10}}&\frac{\sqrt{7}}{5} \\ -\frac{2\sqrt{2}}{5}&-\frac{23}{30}&\frac{2}{\sqrt{35}}&\frac{\sqrt{2}}{5\sqrt{7}}\\-\frac{\sqrt{7}}{\sqrt{10}}&-\frac{2}{\sqrt{35}}&-\frac{3}{14}&-\frac{4\sqrt{2}}{7\sqrt{5}}\\-\frac{\sqrt{7}}{5}
&\frac{\sqrt{2}}{5\sqrt{7}}&-\frac{4\sqrt{2}}{7\sqrt{5}}&\frac{6}{35}\end{array}\right)$& $\mathcal{O}_{20}[2]=\left(\begin{array}{cccc} \frac{2}{27}&0&-\frac{\sqrt{2}}{27\sqrt{35}}&\frac{2}{27\sqrt{7}} \\ 0&\frac{1}{45}&0&\frac{\sqrt{2}}{15\sqrt{7}}\\-\frac{\sqrt{2}}{27\sqrt{35}}&0&\frac{29}{189}&\frac{5\sqrt{10}}{189}\\-\frac{2}{27\sqrt{7}}
&\frac{\sqrt{2}}{15\sqrt{7}}&-\frac{5\sqrt{10}}{189}&-\frac{28}{135}\end{array}\right)$ \\
$\mathcal{O}_{20}^{\prime}[2]=\left(\begin{array}{cccc} \frac{2}{27}&0&-\frac{\sqrt{2}}{27\sqrt{35}}&-\frac{2}{27\sqrt{7}} \\ 0&\frac{1}{45}&0&\frac{\sqrt{2}}{15\sqrt{7}}\\-\frac{\sqrt{2}}{27\sqrt{35}}&0&\frac{29}{189}&-\frac{5\sqrt{10}}{189}\\\frac{2}{27\sqrt{7}}
&\frac{\sqrt{2}}{15\sqrt{7}}&\frac{5\sqrt{10}}{189}&-\frac{28}{135}\end{array}\right)$&$\mathcal{O}_{21}[2]=\left(\begin{array}{cccc} 0&0&-\frac{\sqrt{7}}{54\sqrt{10}}&\frac{\sqrt{7}}{54}\\ 0&-\frac{7}{180}&0&-\frac{1}{15\sqrt{14}}\\-\frac{\sqrt{7}}{54\sqrt{10}}&0&-\frac{1}{252}&\frac{2\sqrt{10}}{189}\\-\frac{\sqrt{7}}{54}&-\frac{1}{15\sqrt{14}}&-\frac{2\sqrt{10}}{189}&\frac{1}{135}\end{array}\right)$ &$\mathcal{O}_{21}^{\prime}[2]=\left(\begin{array}{cccc} 0&0&-\frac{\sqrt{7}}{54\sqrt{10}}&-\frac{\sqrt{7}}{54}\\ 0&-\frac{7}{180}&0&-\frac{1}{15\sqrt{14}}\\-\frac{\sqrt{7}}{54\sqrt{10}}&0&-\frac{1}{252}&-\frac{2\sqrt{10}}{189}\\\frac{\sqrt{7}}{54}&-\frac{1}{15\sqrt{14}}&\frac{2\sqrt{10}}{189}&\frac{1}{135}\end{array}\right)$\\
$\begin{array}{l} \mathcal{O}_{14}^{(\prime)}[3]=\rm {diag}(1,1,1,1) \\ \mathcal{O}_{15}^{(\prime)}[3]=\rm {diag}(-1,\frac{3}{2},\frac{1}{2},-1)\\\mathcal{O}_{17}^{(\prime)}[3]=\rm {diag}(1,\frac{1}{6},\frac{1}{2},1)\\\mathcal{O}_{19}^{(\prime)}[3]=\rm {diag}(-\frac{1}{27},\frac{1}{18},\frac{5}{54},-\frac{1}{27})\end{array}$&$\mathcal{O}_{16}[3]=\left(\begin{array}{cccc} 0&\frac{3}{5\sqrt{2}}&-\frac{1}{\sqrt{5}}&-\frac{4\sqrt{3}}{5} \\ \frac{3}{5\sqrt{2}}&-\frac{3}{35}&-\frac{6\sqrt{2}}{7\sqrt{5}}&-\frac{6\sqrt{6}}{35}\\-\frac{1}{\sqrt{5}}&-\frac{6\sqrt{2}}{7\sqrt{5}}&-\frac{4}{7}&\frac{\sqrt{3}}{7\sqrt{5}}\\-\frac{4\sqrt{3}}{5}
&-\frac{6\sqrt{6}}{35}&\frac{\sqrt{3}}{7\sqrt{5}}&-\frac{22}{35}\end{array}\right)$&$\mathcal{O}_{16}^{\prime}[3]=\left(\begin{array}{cccc} 0&\frac{3}{5\sqrt{2}}&\frac{1}{\sqrt{5}}&-\frac{4\sqrt{3}}{5} \\ \frac{3}{5\sqrt{2}}&-\frac{3}{35}&\frac{6\sqrt{2}}{7\sqrt{5}}&-\frac{6\sqrt{6}}{35}\\\frac{1}{\sqrt{5}}&\frac{6\sqrt{2}}{7\sqrt{5}}&-\frac{4}{7}&-\frac{\sqrt{3}}{7\sqrt{5}}\\-\frac{4\sqrt{3}}{5}
&-\frac{6\sqrt{6}}{35}&-\frac{\sqrt{3}}{7\sqrt{5}}&-\frac{22}{35}\end{array}\right)$\\
$\mathcal{O}_{18}[3]=\left(\begin{array}{cccc} 0&-\frac{1}{5\sqrt{2}}&\frac{1}{\sqrt{5}}&-\frac{2\sqrt{3}}{5} \\ -\frac{1}{5\sqrt{2}}&\frac{23}{105}&\frac{4\sqrt{2}}{7\sqrt{5}}&\frac{2\sqrt{6}}{35}\\\frac{1}{\sqrt{5}}&-\frac{4\sqrt{2}}{7\sqrt{5}}&-\frac{4}{7}&-\frac{\sqrt{3}}{7\sqrt{5}}\\-\frac{2\sqrt{3}}{5}
&\frac{2\sqrt{6}}{35}&\frac{\sqrt{3}}{7\sqrt{5}}&-\frac{11}{35}\end{array}\right)$& $\mathcal{O}_{18}^{\prime}[3]=\left(\begin{array}{cccc} 0&-\frac{1}{5\sqrt{2}}&-\frac{1}{\sqrt{5}}&-\frac{2\sqrt{3}}{5} \\ -\frac{1}{5\sqrt{2}}&\frac{23}{105}&-\frac{4\sqrt{2}}{7\sqrt{5}}&\frac{2\sqrt{6}}{35}\\-\frac{1}{\sqrt{5}}&\frac{4\sqrt{2}}{7\sqrt{5}}&-\frac{4}{7}&\frac{\sqrt{3}}{7\sqrt{5}}\\-\frac{2\sqrt{3}}{5}
&\frac{2\sqrt{6}}{35}&-\frac{\sqrt{3}}{7\sqrt{5}}&-\frac{11}{35}\end{array}\right)$&$\mathcal{O}_{20}[3]=\left(\begin{array}{cccc} -\frac{4}{135}&\frac{\sqrt{2}}{105}&\frac{2\sqrt{5}}{189}&-\frac{4}{315\sqrt{3}} \\ \frac{\sqrt{2}}{105}&\frac{16}{315}&0&-\frac{\sqrt{2}}{35\sqrt{3}}\\-\frac{2\sqrt{5}}{189}&0&\frac{1}{21}&-\frac{\sqrt{5}}{63\sqrt{3}}\\-\frac{4}{315\sqrt{3}}
&-\frac{\sqrt{2}}{35\sqrt{3}}&\frac{\sqrt{5}}{63\sqrt{3}}&\frac{82}{945}\end{array}\right)$\\
$\mathcal{O}_{20}^{\prime}[3]=\left(\begin{array}{cccc} -\frac{4}{135}&\frac{\sqrt{2}}{105}&-\frac{2\sqrt{5}}{189}&-\frac{4}{315\sqrt{3}} \\ \frac{\sqrt{2}}{105}&\frac{16}{315}&0&-\frac{\sqrt{2}}{35\sqrt{3}}\\\frac{2\sqrt{5}}{189}&0&\frac{1}{21}&\frac{\sqrt{5}}{63\sqrt{3}}\\-\frac{4}{315\sqrt{3}}
&-\frac{\sqrt{2}}{35\sqrt{3}}&-\frac{\sqrt{5}}{63\sqrt{3}}&\frac{82}{945}\end{array}\right)$&$\mathcal{O}_{21}[3]=\left(\begin{array}{cccc} 0&\frac{1}{30\sqrt{2}}&\frac{\sqrt{5}}{54}&-\frac{1}{45\sqrt{3}}\\ \frac{1}{30\sqrt{2}}&\frac{1}{90}&0&-\frac{\sqrt{2}}{35\sqrt{3}}\\-\frac{\sqrt{5}}{54}&0&-\frac{2}{189}&\frac{\sqrt{5}}{126\sqrt{3}}\\-\frac{1}{45\sqrt{3}}&-\frac{\sqrt{2}}{35\sqrt{3}}&-\frac{\sqrt{5}}{126\sqrt{3}}&-\frac{11}{1890}\end{array}\right)$
&$\mathcal{O}_{21}^{\prime}[3]=\left(\begin{array}{cccc} 0&\frac{1}{30\sqrt{2}}&-\frac{\sqrt{5}}{54}&-\frac{1}{45\sqrt{3}}\\ \frac{1}{30\sqrt{2}}&\frac{1}{90}&0&-\frac{\sqrt{2}}{35\sqrt{3}}\\\frac{\sqrt{5}}{54}&0&-\frac{2}{189}&-\frac{\sqrt{5}}{126\sqrt{3}}\\-\frac{1}{45\sqrt{3}}&-\frac{\sqrt{2}}{35\sqrt{3}}&\frac{\sqrt{5}}{126\sqrt{3}}&-\frac{11}{1890}\end{array}\right)$      \\
\bottomrule[1pt]\bottomrule[1pt]
\end{tabular}
\end{table*}


\begin{thebibliography}{99}
\bibitem{GellMann:1964nj}
  M.~Gell-Mann,
  A Schematic Model of Baryons and Mesons,
  \href{https://www.sciencedirect.com/science/article/abs/pii/S0031916364920013?via\%3Dihub}{Phys.\ Lett.\  {\bf 8}, 214 (1964)}.

\bibitem{Zweig:1981pd}
G.~Zweig,
An SU(3) model for strong interaction symmetry and its breaking. Version 1,
\href{http://cds.cern.ch/record/352337}{CERN-TH-401.}

\bibitem{Chen:2016qju}
  H.~X.~Chen, W.~Chen, X.~Liu, and S.~L.~Zhu,
  The hidden-charm pentaquark and tetraquark states,
  \href{http://linkinghub.elsevier.com/retrieve/pii/S037015731630103X}{Phys.\ Rep.\  {\bf 639}, 1 (2016)}.

\bibitem{Liu:2019zoy}
  Y.~R.~Liu, H.~X.~Chen, W.~Chen, X.~Liu, and S.~L.~Zhu,
  Pentaquark and tetraquark states,
  \href{https://www.sciencedirect.com/science/article/pii/S0146641019300304?via\%3Dihub}{Prog.\ Part.\ Nucl.\ Phys.\  {\bf 107}, 237 (2019)}.

\bibitem{Choi:2003ue}
  S.~K.~Choi {\it et al.} [Belle Collaboration],
  Observation of a Narrow Charmonium-Like State in Exclusive $B^{\pm} \to K^{\pm} \pi^+ \pi^- J/\psi$ Decays,
  \href{https://journals.aps.org/prl/abstract/10.1103/PhysRevLett.91.262001}{Phys.\ Rev.\ Lett.\  {\bf 91}, 262001 (2003)}.

\bibitem{Olsen:2017bmm}
  S.~L.~Olsen, T.~Skwarnicki, and D.~Zieminska,
  Nonstandard heavy mesons and baryons: Experimental evidence,
  \href{https://journals.aps.org/rmp/abstract/10.1103/RevModPhys.90.015003}{Rev.\ Mod.\ Phys.\  {\bf 90}, 015003 (2018)}.

\bibitem{Guo:2017jvc}
  F.~K.~Guo, C.~Hanhart, U.~G.~Mei$\ss$ner, Q.~Wang, Q.~Zhao, and B.~S.~Zou,
  Hadronic molecules,
  \href{https://journals.aps.org/rmp/abstract/10.1103/RevModPhys.90.015004}{Rev.\ Mod.\ Phys.\  {\bf 90}, 015004 (2018)}.

\bibitem{Liu:2013waa}
  X.~Liu,
  An overview of $XYZ$ new particles,
  \href{http://dx.doi.org/10.1007/s11434-014-0407-2}{Chin.\ Sci.\ Bull.\  {\bf 59}, 3815 (2014)}.

\bibitem{Hosaka:2016pey}
  A.~Hosaka, T.~Iijima, K.~Miyabayashi, Y.~Sakai, and S.~Yasui,
  Exotic hadrons with heavy flavors: $X$, $Y$, $Z$, and related states,
  \href{http://dx.doi.org/10.1093/ptep/ptw045}{Prog. Theor. Exp. Phys. {\bf 2016}, 062C01 (2016)}.

\bibitem{Brambilla:2019esw}
N.~Brambilla, S.~Eidelman, C.~Hanhart, A.~Nefediev, C.~P.~Shen, C.~E.~Thomas, A.~Vairo, and C.~Z.~Yuan,
The $XYZ$ states: Experimental and theoretical status and perspectives,
\href{https://www.sciencedirect.com/science/article/pii/S0370157320301915?via\%3Dihub}{Phys. Rep. \textbf{873}, 1 (2020)}.

\bibitem{Tcc:talk}
See the talk by Franz Muheim at the European Physical Society
conference on high energy physics 2021 on July 28,
\href{https://indico.desy.de/event/28202/contributions/102717/}{https://indico.desy.de/event/28202/contributions/102717/}.

\bibitem{Li:2012ss}
N.~Li, Z.~F.~Sun, X.~Liu and S.~L.~Zhu, Coupled-channel analysis of
The possible $D^{(*)}D^{(*)}, \overline{B}^{(*)}\overline{B}^{(*)}$ and $D^{(*)}\overline{B}^{(*)}$ molecular states,
\href{https://journals.aps.org/prd/abstract/10.1103/PhysRevD.88.114008}{Phys. Rev. D \textbf{88}, no.11, 114008 (2013)}.

\bibitem{Xu:2017tsr}
H.~Xu, B.~Wang, Z.~W.~Liu and X.~Liu, $D D^{*}$ potentials in chiral perturbation theory and possible molecular states,
\href{https://journals.aps.org/prd/abstract/10.1103/PhysRevD.99.014027}{Phys. Rev. D \textbf{99}, no.1, 014027 (2019)}.

\bibitem{Liu:2019stu}
M.~Z.~Liu, T.~W.~Wu, M.~Pavon Valderrama, J.~J.~Xie and L.~S.~Geng,
Heavy-quark spin and flavor symmetry partners of the $X(3872)$ revisited: What can we learn from the one boson exchange model?,
\href{https://journals.aps.org/prd/abstract/10.1103/PhysRevD.98.114030}{Phys. Rev. D \textbf{99}, no.9, 094018 (2019)}.

\bibitem{Li:2021zbw}
N.~Li, Z.~F.~Sun, X.~Liu and S.~L.~Zhu,
Perfect $DD^*$ molecular prediction matching the $T_{cc}$ observation at LHCb,
\href{https://arxiv.org/abs/2107.13748}{arXiv:2107.13748}.

\bibitem{Chen:2021vhg}
R.~Chen, Q.~Huang, X.~Liu and S.~L.~Zhu,
Another doubly charmed molecular resonance $T_{cc}^{\prime+}(3876)$,
\href{https://arxiv.org/abs/2108.01911}{arXiv:2108.01911}.

\bibitem{Wong:2003xk}
  C.~Y.~Wong,
  Molecular states of heavy quark mesons,
  \href{https://journals.aps.org/prc/abstract/10.1103/PhysRevC.69.055202}{Phys.\ Rev.\ C {\bf 69}, 055202 (2004)}.


\bibitem{Swanson:2003tb}
  E.~S.~Swanson,
  Short range structure in the $X(3872)$,
  \href{https://www.sciencedirect.com/science/article/pii/S0370269304004599?via\%3Dihub}{Phys.\ Lett.\ B {\bf 588}, 189 (2004)}.

\bibitem{Suzuki:2005ha}
  M.~Suzuki,
  The $X(3872)$ boson: Molecule or charmonium,
  \href{https://journals.aps.org/prd/abstract/10.1103/PhysRevD.72.114013}{Phys.\ Rev.\ D {\bf 72}, 114013 (2005)}.

\bibitem{Liu:2008fh}
  Y.~R.~Liu, X.~Liu, W.~Z.~Deng, and S.~L.~Zhu,
  Is $X(3872)$ really a molecular state?,
  \href{https://link.springer.com/article/10.1140/epjc/s10052-008-0640-4}{Eur.\ Phys.\ J.\ C {\bf 56}, 63 (2008)}.

\bibitem{Thomas:2008ja}
  C.~E.~Thomas and F.~E.~Close,
  Is $X(3872)$ a molecule?,
  \href{https://journals.aps.org/prd/abstract/10.1103/PhysRevD.78.034007}{Phys.\ Rev.\ D {\bf 78}, 034007 (2008)}.

\bibitem{Liu:2008tn}
  X.~Liu, Z.~G.~Luo, Y.~R.~Liu, and S.~L.~Zhu,
  $X(3872)$ and other possible heavy molecular states,
  \href{https://link.springer.com/article/10.1140\%2Fepjc\%2Fs10052-009-1020-4}{Eur.\ Phys.\ J.\ C {\bf 61}, 411 (2009)}.

\bibitem{Lee:2009hy}
  I.~W.~Lee, A.~Faessler, T.~Gutsche, and V.~E.~Lyubovitskij,
  $X(3872)$ as a molecular $DD^*$ state in a potential model,
  \href{https://journals.aps.org/prd/abstract/10.1103/PhysRevD.80.094005}{Phys.\ Rev.\ D {\bf 80}, 094005 (2009)}.

\bibitem{Zhao:2014gqa}
L.~Zhao, L.~Ma, and S.~L.~Zhu,
Spin-orbit force, recoil corrections, and possible $B \bar{B}^{*}$ and $D \bar{D}^{*}$  molecular states,
\href{https://journals.aps.org/prd/abstract/10.1103/PhysRevD.89.094026}{Phys. Rev. D \textbf{89}, 094026 (2014)}.

\bibitem{Li:2012cs}
N.~Li and S.~L.~Zhu,
Isospin breaking, coupled-channel effects and diagnosis of $X(3872)$,
\href{https://journals.aps.org/prd/abstract/10.1103/PhysRevD.86.074022}{Phys. Rev. D \textbf{86}, 074022 (2012)}.

\bibitem{Voloshin:2003nt}
  M.~B.~Voloshin,
  Interference and binding effects in decays of possible molecular component of $X(3872)$,
  \href{http://dx.doi.org/10.1016/j.physletb.2003.11.014}{Phys.\ Lett.\ B {\bf 579}, 316 (2004)}.

\bibitem{Close:2003sg}
  F.~E.~Close and P.~R.~Page,
   The $D^{*0} \bar D^0$ threshold resonance,
  \href{http://dx.doi.org/10.1016/j.physletb.2003.10.032}{Phys.\ Lett.\ B {\bf 578}, 119 (2004)}.

\bibitem{Tornqvist:2004qy}
  N.~A.~Tornqvist,
   Isospin breaking of the narrow charmonium state of Belle at 3872 MeV as a deuson,
  \href{http://dx.doi.org/10.1016/j.physletb.2004.03.077}{Phys.\ Lett.\ B {\bf 590}, 209 (2004)}.

\bibitem{He:2014nya}
  J.~He,
  Study of the $B\bar{B}^*/D\bar{D}^*$ bound states in a Bethe-Salpeter approach,
  \href{https://journals.aps.org/prd/abstract/10.1103/PhysRevD.90.076008}{Phys.\ Rev.\ D {\bf 90}, 076008 (2014).}

\bibitem{Choi:2007wga}
  S.~K.~Choi {\it et al.} [Belle Collaboration],
  Observation of a ResonanceLike Structure in the $\pi^\pm \psi^\prime$ Mass Distribution in Exclusive $B \to K \pi^\pm \psi^\prime$ Decays,
  \href{https://journals.aps.org/prl/abstract/10.1103/PhysRevLett.100.142001}{Phys.\ Rev.\ Lett.\  {\bf 100}, 142001 (2008)}.

\bibitem{Liu:2007bf}
  X.~Liu, Y.~R.~Liu, W.~Z.~Deng and S.~L.~Zhu,
  Is $Z^+(4430)$ a loosely bound molecular state?,
  \href{https://journals.aps.org/prd/abstract/10.1103/PhysRevD.77.034003}{Phys.\ Rev.\ D {\bf 77}, 034003 (2008)}.

\bibitem{Liu:2008xz}
  X.~Liu, Y.~R.~Liu, W.~Z.~Deng and S.~L.~Zhu,
  $Z^+(4430)$ as a $D_1^\prime D^* (D_1 D^*)$ molecular state,
  \href{https://journals.aps.org/prd/abstract/10.1103/PhysRevD.77.094015}{Phys.\ Rev.\ D {\bf 77}, 094015 (2008)}.

\bibitem{Aubert:2005rm}
  B.~Aubert {\it et al.} [BaBar Collaboration],
  Observation of a Broad Structure in the $\pi^+ \pi^- J/\psi$ Mass Spectrum Around 4.26-GeV/c$^2$,
  \href{https://journals.aps.org/prl/abstract/10.1103/PhysRevLett.95.142001}{Phys.\ Rev.\ Lett.\  {\bf 95}, 142001 (2005)}.

\bibitem{Ding:2008gr}
  G.~J.~Ding,
  Are $Y(4260)$ and {\rm$Z_2^{+}$(4250)} ${\rm D_1D}$ or ${\rm D_0D^{*}}$ hadronic molecules?
  \href{https://journals.aps.org/prd/abstract/10.1103/PhysRevD.79.014001}{Phys.\ Rev.\ D {\bf 79}, 014001 (2009)}.

\bibitem{Cleven:2013mka}
  M.~Cleven, Q.~Wang, F.~K.~Guo, C.~Hanhart, U.~G.~Mei$\beta$ner, and Q.~Zhao,
  $Y(4260)$ as the first $S$-wave open charm vector molecular state?,
  \href{https://journals.aps.org/prd/abstract/10.1103/PhysRevD.90.074039}{Phys.\ Rev.\ D {\bf 90}, 074039 (2014)}.


\bibitem{Wang:2013kra}
  Q.~Wang, M.~Cleven, F.~K.~Guo, C.~Hanhart, U.~G.~Mei$\beta$ner, X.~G.~Wu, and Q.~Zhao,
  $Y(4260)$: Hadronic molecule versus hadro-charmonium interpretation,
  \href{https://journals.aps.org/prd/abstract/10.1103/PhysRevD.89.034001}{Phys.\ Rev.\ D {\bf 89}, 034001 (2014)}.

\bibitem{BaBar:2006ait}
B.~Aubert \textit{et al.} [BaBar Collaboration],
Evidence of a broad structure at an invariant mass of 4.32- $GeV/c^{2}$ in the reaction $e^{+} e^{-} \to \pi^{+} \pi^{-} \psi_{2S}$ measured at BaBar,
\href{https://journals.aps.org/prl/abstract/10.1103/PhysRevLett.98.212001}{Phys. Rev. Lett. \textbf{98}, 212001 (2007)}.

\bibitem{Close:2009ag}
F.~Close and C.~Downum,
On the possibility of Deeply Bound Hadronic Molecules from single Pion Exchange,
\href{https://journals.aps.org/prl/abstract/10.1103/PhysRevLett.102.242003}{Phys. Rev. Lett. \textbf{102}, 242003 (2009)}.

\bibitem{Close:2010wq}
F.~Close, C.~Downum and C.~E.~Thomas,
Novel Charmonium and Bottomonium Spectroscopies due to Deeply Bound Hadronic Molecules from Single Pion Exchange,
\href{https://journals.aps.org/prd/abstract/10.1103/PhysRevD.81.074033}{Phys. Rev. D \textbf{81}, 074033 (2010)}.

\bibitem{Dong:2021bvy}
X.~K.~Dong, F.~K.~Guo and B.~S.~Zou,
A survey of heavy-heavy hadronic molecules,
\href{https://arxiv.org/abs/2108.02673}{arXiv:2108.02673}.

\bibitem{Bediaga:2018lhg}
R.~Aaij \textit{et al.} [LHCb Collaboration],
Physics case for an LHCb Upgrade II-Opportunities in flavour physics, and beyond, in the HL-LHC era,
\href{https://arxiv.org/abs/1808.08865}{arXiv:1808.08865}.

\bibitem{Wang:2021hql}
F.~L.~Wang, X.~D.~Yang, R.~Chen and X.~Liu,
Hidden-charm pentaquarks with triple strangeness due to the $\Omega_{c}^{(*)}\bar{D}_s^{(*)}$ interactions,
\href{https://journals.aps.org/prd/abstract/10.1103/PhysRevD.103.054025}{Phys. Rev. D \textbf{103}, 054025 (2021)}.

\bibitem{Wang:2019nwt}
  F.~L.~Wang, R.~Chen, Z.~W.~Liu, and X.~Liu,
  Probing new types of $P_c$ states inspired by the interaction between $S-$wave charmed baryon and anti-charmed meson in a $\bar T$ doublet,
  \href{https://journals.aps.org/prc/abstract/10.1103/PhysRevC.101.025201}{Phys.\ Rev.\ C {\bf 101},  025201 (2020)}.

\bibitem{Wang:2019aoc}
  F.~L.~Wang, R.~Chen, Z.~W.~Liu, and X.~Liu,
  Possible triple-charm molecular pentaquarks from $\Xi_{cc}D_1/\Xi_{cc}D_2^*$ interactions,
  \href{https://journals.aps.org/prd/abstract/10.1103/PhysRevD.99.054021}{Phys.\ Rev.\ D {\bf 99}, 054021 (2019)}.

\bibitem{Wang:2020dya}
F.~L.~Wang and X.~Liu,
Exotic double-charm molecular states with hidden or open strangeness and around $4.5\sim 4.7$ GeV,
\href{https://journals.aps.org/prd/abstract/10.1103/PhysRevD.102.094006}{Phys. Rev. D \textbf{102}, 094006 (2020)}.

\bibitem{Wang:2020bjt}
F.~L.~Wang, R.~Chen, and X.~Liu,
Prediction of hidden-charm pentaquarks with double strangeness,
\href{https://journals.aps.org/prd/abstract/10.1103/PhysRevD.103.034014}{Phys. Rev. D \textbf{103}, 034014 (2021)}.

\bibitem{Chen:2018pzd}
  R.~Chen, F.~L.~Wang, A.~Hosaka and X.~Liu,
  Exotic triple-charm deuteronlike hexaquarks,
  \href{https://journals.aps.org/prd/abstract/10.1103/PhysRevD.97.114011}{Phys.\ Rev.\ D {\bf 97}, 114011 (2018)}.

\bibitem{Wang:2021aql}
F.~L.~Wang, X.~D.~Yang, R.~Chen and X.~Liu,
Correlation of the hidden-charm molecular tetraquarks and the charmonium-like structures existing in the $B\to XYZ+K$,
\href{https://arxiv.org/abs/2103.04698}{arXiv:2103.04698}.

\bibitem{Casalbuoni:1992gi}
  R.~Casalbuoni, A.~Deandrea, N.~Di Bartolomeo, R.~Gatto, F.~Feruglio, and G.~Nardulli,
  Light vector resonances in the effective chiral Lagrangian for heavy mesons,
  \href{https://www.sciencedirect.com/science/article/abs/pii/037026939291189G}{Phys.\ Lett.\ B {\bf 292}, 371 (1992)}.

\bibitem{Casalbuoni:1996pg}
  R.~Casalbuoni, A.~Deandrea, N.~Di Bartolomeo, R.~Gatto, F.~Feruglio, and G.~Nardulli,
  Phenomenology of heavy meson chiral Lagrangians,
  \href{https://www.sciencedirect.com/science/article/pii/S0370157396000270}{Phys.\ Rep.\  {\bf 281}, 145 (1997)}.

\bibitem{Yan:1992gz}
  T.~M.~Yan, H.~Y.~Cheng, C.~Y.~Cheung, G.~L.~Lin, Y.~C.~Lin, and H.~L.~Yu,
  Heavy quark symmetry and chiral dynamics,
  \href{https://journals.aps.org/prd/abstract/10.1103/PhysRevD.46.1148}{Phys.\ Rev.\ D {\bf 46}, 1148 (1992);}
   \href{https://journals.aps.org/prd/abstract/10.1103/PhysRevD.55.5851}{[Phys.\ Rev.\ D {\bf 55}, 5851E (1997)]}.

\bibitem{Bando:1987br}
  M.~Bando, T.~Kugo, and K.~Yamawaki,
  Nonlinear realization and hidden local symmetries,
  \href{https://www.sciencedirect.com/science/article/abs/pii/0370157388900191?via\%3Dihub}{Phys.\ Rep.\  {\bf 164}, 217 (1988)}.

\bibitem{Harada:2003jx}
  M.~Harada and K.~Yamawaki,
  Hidden local symmetry at loop: A new perspective of composite gauge boson and chiral phase transition,
  \href{https://www.sciencedirect.com/science/article/abs/pii/S037015730300139X?via\%3Dihub}{Phys.\ Rep.\  {\bf 381}, 1 (2003)}.

\bibitem{Wise:1992hn}
  M.~B.~Wise,
  Chiral perturbation theory for hadrons containing a heavy quark,
   \href{https://journals.aps.org/prd/abstract/10.1103/PhysRevD.45.R2188}{Phys.\ Rev.\ D {\bf 45}, R2188 (1992)}.

\bibitem{Falk:1992cx}
A.~F.~Falk and M.~E.~Luke, Strong decays of excited heavy mesons in chiral perturbation theory,
\href{https://www.sciencedirect.com/science/article/abs/pii/037026939290618E?via\%3Dihub}{Phys.\ Lett.\  B {\bf 292}, 119 (1992)}.

\bibitem{Isola:2003fh}
  C.~Isola, M.~Ladisa, G.~Nardulli, and P.~Santorelli,
  Charming penguins in $B\to K^{*}\pi, K(\rho,\omega,\phi)$ decays,
  \href{https://journals.aps.org/prd/abstract/10.1103/PhysRevD.68.114001}{Phys.\ Rev.\ D {\bf 68}, 114001 (2003)}.

\bibitem{Cleven:2016qbn}
  M.~Cleven and Q.~Zhao,
  Cross section line shape of $e^+e^-\to\chi_{c0}\omega$ around the $Y(4260)$ mass region,
 \href{https://linkinghub.elsevier.com/retrieve/pii/S0370269317301545}{Phys.\ Lett.\ B {\bf 768}, 52 (2017)}.

\bibitem{Dong:2019ofp}
X.~K.~Dong, Y.~H.~Lin, and B.~S.~Zou,
Prediction of an exotic state around 4240 MeV with $J^{PC}=1^{-+}$ as C-parity partner of $Y(4260)$ in molecular picture,
\href{https://journals.aps.org/prd/abstract/10.1103/PhysRevD.101.076003}{Phys. Rev. D \textbf{101},  076003 (2020)}.

\bibitem{He:2019csk}
J.~He, Y.~Liu, J.~T.~Zhu, and D.~Y.~Chen,
$Y(4626)$ as a molecular state from interaction ${D}^*_s{\bar{D}}_{s1}(2536)-{D}_s{\bar{D}}_{s1}(2536)$,
\href{https://link.springer.com/article/10.1140/epjc/s10052-020-7820-2}{Eur. Phys. J. C \textbf{80},  246 (2020)}.

\bibitem{Wang:2020lua}
Z.~Y.~Wang, J.~J.~Qi, J.~Xu, and X.~H.~Guo,
Studying the $D_1D$ molecule in the Bethe-Salpeter equation approach,
\href{https://journals.aps.org/prd/abstract/10.1103/PhysRevD.102.036008}{Phys. Rev. D \textbf{102}, 036008 (2020)}.

\bibitem{Riska:2000gd}
  D.~O.~Riska and G.~E.~Brown,
  Nucleon resonance transition couplings to vector mesons,
  \href{https://www.sciencedirect.com/science/article/pii/S0375947400003626}{Nucl.\ Phys.\ {\bf A} {\bf 679}, 577 (2001)}.

\bibitem{Zyla:2020zbs}
 P.~A.~Zyla {\it et al.} [Particle Data Group],
 Review of Particle Physics,
 \href{https://academic.oup.com/ptep/article/2020/8/083C01/5891211}{PTEP \textbf{2020}, 083C01 (2020)}.

\bibitem{Cheng:2010yd}
  H.~Y.~Cheng and K.~C.~Yang,
  Charmless hadronic $B$ decays into a tensor meson,
  \href{https://journals.aps.org/prd/abstract/10.1103/PhysRevD.83.034001}{Phys.\ Rev.\ D {\bf 83}, 034001 (2011)}.

\bibitem{Breit:1929zz}
  G.~Breit,
  The effect of retardation on the interaction of two electrons,
  \href{https://journals.aps.org/pr/abstract/10.1103/PhysRev.34.553}{Phys.\ Rev.\  {\bf 34}, 553 (1929)}.

\bibitem{Breit:1930zza}
  G.~Breit,
  The fine structure of HE as a test of the spin interactions of two electrons,
  \href{https://journals.aps.org/pr/abstract/10.1103/PhysRev.36.383}{Phys.\ Rev.\  {\bf 36}, 383 (1930)}.


\bibitem{Tornqvist:1993ng}
  N.~A.~Tornqvist,
  From the deuteron to deusons, an analysis of deuteronlike meson-meson bound states,
   \href{https://link.springer.com/article/10.1007\%2FBF01413192}{Z.\ Phys.\ C {\bf 61}, 525 (1994)}.

\bibitem{Tornqvist:1993vu}
  N.~A.~Tornqvist,
  On deusons or deuteron-like meson-meson bound states,
  \href{https://link.springer.com/article/10.1007\%2FBF02734018}{Nuovo Cimento Soc. Ital. Fis.  {\bf 107A}, 2471 (1994)}.

\bibitem{Chen:2017jjn}
  R.~Chen, A.~Hosaka, and X.~Liu,
  Prediction of triple-charm molecular pentaquarks,
  \href{https://journals.aps.org/prd/abstract/10.1103/PhysRevD.96.114030}{Phys.\ Rev.\ D {\bf 96}, 114030 (2017)}.

\bibitem{Chen:2017xat}
  R.~Chen, A.~Hosaka, and X.~Liu,
  Searching for possible $\Omega_c$-like molecular states from meson-baryon interaction,
  \href{https://journals.aps.org/prd/abstract/10.1103/PhysRevD.97.036016}{Phys.\ Rev.\ D {\bf 97}, 036016 (2018)}.

\bibitem{Li:2012bt}
  N.~Li and S.~L.~Zhu,
  Hadronic Molecular States Composed of Heavy Flavor Baryons,
  \href{https://journals.aps.org/prd/abstract/10.1103/PhysRevD.86.014020}{Phys.\ Rev.\ D {\bf 86}, 014020 (2012)}.

\bibitem{Chen:2015add}
R.~Chen, X.~Liu, Y.~R.~Liu, and S.~L.~Zhu,
Predictions of the hidden-charm molecular states with four-quark component,
\href{https://link.springer.com/article/10.1140/epjc/s10052-016-4166-x}{Eur. Phys. J. C \textbf{76}, 319 (2016)}.

\end{thebibliography}
\end{document}